\documentclass[sigconf,9pt]{acmart}
\renewcommand\footnotetextcopyrightpermission[1]{} 
\usepackage{caption,subcaption}
\usepackage{cleveref}

\crefformat{section}{\S#2#1#3} 
\crefformat{subsection}{\S#2#1#3}
\crefformat{subsubsection}{\S#2#1#3}

\usepackage{amsmath}
\usepackage{algorithm}
\usepackage{algpseudocode}

\author{Cheng Jiang, Yihe Yan, Yanxiang Wang, Jiawei Hu, Chun Tung Chou, Wen Hu}
\affiliation{
    \institution{
    School of Computer Science and Engineering, 
    University of New South Wales}
    \city{Sydney}
    \country{Australia}
}
\email{{cheng.jiang1, yihe.yan, yanxiang.wang, jiawei.hu, c.t.chou, wen.hu}@unsw.edu.au}

\AtBeginDocument{%
  \providecommand\BibTeX{{%
    \normalfont B\kern-0.5em{\scshape i\kern-0.25em b}\kern-0.8em\TeX}}}

\definecolor{re}{RGB}{0,0,0}
\definecolor{rectc}{RGB}{0,0,0}

\begin{document}

\title{CARTS: Cooperative and Adaptive Resource Triggering and Stitching for 5G ISAC}

\begin{abstract}
\noindent
This paper presents \textbf{CARTS}, an adaptive 5G uplink sensing scheme designed to provide \textit{Integrated Sensing and Communication} (ISAC) services. The performance of both communication and sensing fundamentally depends on the availability of accurate and up-to-date \textit{channel state information} (CSI). 
In modern 5G networks, uplink CSI is derived from two reference signals: the \textit{demodulation reference signal} (DMRS) and the \textit{sounding reference signal} (SRS). However, current base station implementations treat these CSI measurements as separate information streams. The key innovation of CARTS is to \textit{fuse} these two CSI streams, thereby increasing the frequency of CSI updates and extending sensing opportunities to more users.
CARTS addresses two key challenges: (i) a novel channel stitching and compensation method that integrates asynchronous CSI estimates from DMRS and SRS, despite their different time and frequency allocations, and (ii) a real-time SRS triggering algorithm that complements the inherently uncontrollable DMRS schedule, ensuring sufficient and non-redundant sensing opportunities for all users.
\textcolor{re}{Our trace-driven evaluation shows that CARTS significantly improves scalability, achieving a channel estimation error (NMSE) of 0.167 and UE tracking accuracy of 85 cm while supporting twice the number of users as a periodic SRS-only baseline with similar performance. By opportunistically combining DMRS and SRS, CARTS therefore provides a practical, standard-compliant solution to improve CSI availability for ISAC without requiring additional radio resources.}

\end{abstract}

\acmSubmissionID{39}
\maketitle
\section{INTRODUCTION}
Integrated Sensing and Communication (ISAC) has emerged as a key enabler for 5G-Advanced and upcoming 6G wireless systems, addressing the diverse sensing demands of advanced applications such as autonomous driving, digital twins, extended reality (XR), and smart factories~\cite{saad2019vision, liu2022integrated}. By equipping the \textit{gNB} (Next Generation NodeB, the 5G term for a base station) with sensing capabilities, ISAC allows cellular networks to deliver not only high-speed connectivity but also precise sensing and positioning services to users within their coverage.

Previous research on 5G ISAC has primarily focused on developing new waveforms and frame structures to integrate radar sensing functionalities into base stations~\cite{li2024frame, zheng2023inner}. However, these approaches often compromise compatibility with existing 5G devices, leading to substantial upgrade costs given the extensive deployment of 5G infrastructure and \textit{User Equipment} (UE). Fortunately, since both 5G and Wi-Fi utilize \textit{Orthogonal Frequency Division Multiplexing} (OFDM), extensive research on leveraging \textit{channel state information} (CSI) for Wi-Fi sensing~\cite{kotaru2015spotfi, wu2021witraj, qian2017widar} suggests the feasibility of 5G sensing through CSI acquisition. Research indicates that a combination of large bandwidth and frequent CSI measurements enhances sensing accuracy~\cite{jiang2014communicating}. Moreover, timely CSI measurements significantly improve communication throughput by adapting to rapidly changing channel conditions~\cite{bejarano2016resilient}. Consequently, the availability of accurate and timely CSI underpins the performance of both communication and sensing in 5G ISAC. Since CSI is obtained by transmitting pilot symbols over the radio channel, channel measurements consume resources in both the time and frequency domains. This paper focuses on enhancing CSI availability in the uplink channel for more users \textit{without requiring additional radio resources}, ensuring compatibility with the current 5G standard while enabling low-cost deployment within a virtualized Radio Access Network (vRAN).


The 5G standard employs the \textit{Sounding Reference Signal} (SRS) for uplink CSI measurement. Given the limited channel-sounding (sensing) resources in both frequency and time domains, 5G networks must efficiently allocate spectrum and time slots among numerous users. Traditional approaches often rely on fixed or periodic resource allocation strategies that fail to adapt to dynamic network conditions. In response, recent advancements have introduced more adaptive and intelligent scheduling mechanisms. However, these solutions primarily focus on \textsl{either} improving communication efficiency or enhancing sensing accuracy, rather than jointly optimizing both aspects. 

For instance, \textit{TRADER}~\cite{fiandrino2021traffic} utilizes machine learning-based traffic prediction for dynamic SRS scheduling, ensuring timely CSI acquisition for UE engaged in data transmissions, but does not consider sensing requirements. Conversely, \textit{ElaSe}~\cite{chen2024elase} adjusts the interval of periodic SRS based on user motion, reducing trajectory errors but under-allocating resources for static UEs and enforcing a rigid velocity-periodicity relationship, limiting scalability. However, the fundamental challenge remains the scarcity of SRS resources. A scalable scheduling algorithm is needed to jointly address both communication and sensing requirements, as these objectives are not inherently conflicting.


To overcome the sensing limitations of SRS alone, we note that 5G networks can also acquire CSI using the \textit{DeModulation Reference Signal} (DMRS). 
\textcolor{re}{In standard 5G operation, DMRS and SRS serve distinct purposes: demodulation and channel sounding, respectively, and their CSI measurements are typically processed in separate streams. CARTS identifies an opportunity to \textbf{fuse} these streams for enhanced sensing.}

However, a key challenge in this fusion process is \textit{asynchronization}, as DMRS and SRS channel measurements occur at different times and potentially different frequency bands. Simply combining these measurements without compensation could result in inaccurate channel representations, leading to errors in estimated gains, phase shifts, and timing advance (TA) offsets. These errors degrade performance, increasing block error rates and potentially causing connection failures. Additionally, it is crucial to recognize that DMRS measurements always align with their associated uplink traffic in frequency. Given that DMRS scheduling is inherently dictated by uplink data transmission, it remains uncontrollable. Instead, we focus on optimizing SRS allocation to complement DMRS, ensuring rational utilization of radio resources. For instance, avoiding redundant DMRS and SRS measurements in the same frequency band within a short interval prevents resource wastage.


This paper introduces \textbf{CARTS} (\textit{Cooperative and Adaptive Resource Triggering and Stitching}), a framework that addresses these challenges with the following key contributions:

\begin{itemize}
    \item \textbf{Joint DMRS-SRS Sensing Mechanism:} CARTS introduces a joint channel estimation framework that fuses CSI measurements from DMRS and SRS, expanding sensing opportunities beyond what is achievable using SRS alone. This fusion enhances resource sharing among users without requiring additional radio resources.
    
    \item \textbf{Asynchronization Mitigation:} To address the temporal misalignment between DMRS and SRS transmissions, CARTS incorporates a novel channel stitching and compensation method for synchronizing CSI measurements (\cref{sync_comp}).
    
    \item \textbf{Adaptive SRS Triggering:} Recognizing that DMRS-based sensing opportunities are inherently linked to uplink traffic, CARTS features a priority-based adaptive SRS triggering algorithm (\cref{adaptive_algo}).
    This algorithm minimizes spectrum overlap between DMRS and SRS while ensuring sufficient sensing opportunities for each UE based on its communication and sensing demands.
\end{itemize}

\section{PRELIMINARY} \label{prelimnary}

\subsection{Frame Structure}

\begin{figure}[t!]
    \centering
    \includegraphics[width=\columnwidth]{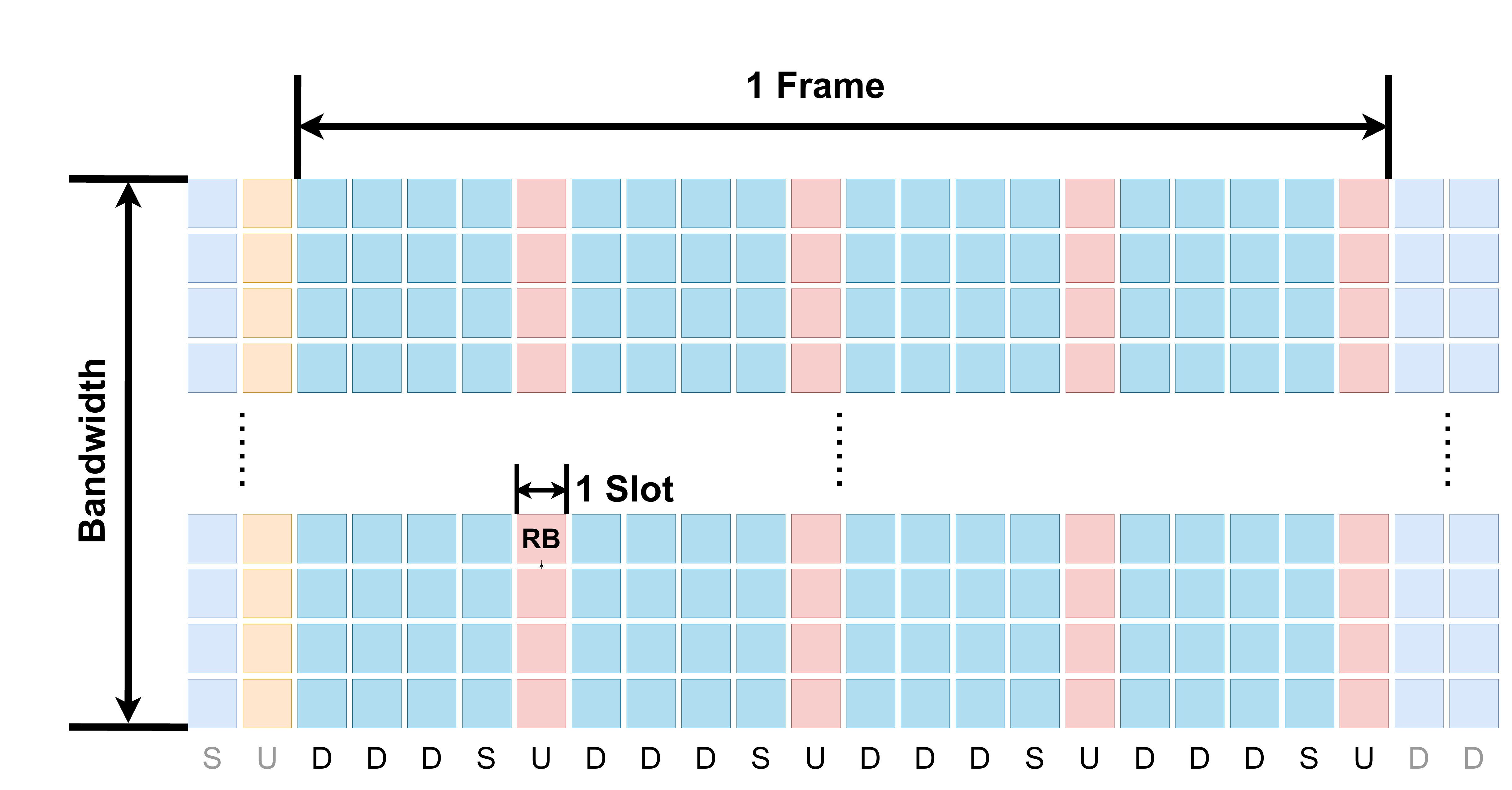}
    \caption{Example 5G PHY frame structure.}
    \label{fig:frame_structure}
\end{figure}

The 5G New Radio (NR) frame structure is a highly flexible and scalable architecture designed to accommodate diverse use cases, including \textit{enhanced mobile broadband} (eMBB), \textit{ultra-reliable low-latency communication} (URLLC), and \textit{massive machine-type communication} (mMTC). This paper focuses on \textit{time-division duplex} (TDD) operation.

A 5G NR frame spans a fixed duration of \textbf{10 milliseconds (ms)} and is divided into \textbf{10 subframes} of 1 ms each. Each subframe is further partitioned into multiple slots, with the number of slots per subframe determined by the chosen numerology, which defines the subcarrier spacing (SCS). The scalable SCS values in 5G NR are 15, 30, 60, 120, and 240 kHz, enabling the system to support varying latency, bandwidth, and frequency range requirements. For instance, numerologies corresponding to SCS values of 15/30/60 kHz result in 1/2/4 slots per subframe, respectively.

In TDD mode, slots are allocated for \textbf{downlink (D)}, \textbf{uplink (U)}, or \textbf{special (S)} transmissions. Special slots contain guard symbols to facilitate the base station’s transition between downlink and uplink transmissions. Fig.~\ref{fig:frame_structure} illustrates an example frame structure where the SCS is configured to 30 kHz, the TDD slot format is set to "DDDSU," and the TDD periodicity is 2.5 ms.

\subsection{Uplink Resource Block and Reference Signals} \label{uplink_ref}

\begin{figure}[t!]
    \centering
    \hspace{0cm}\includegraphics[width=\columnwidth]{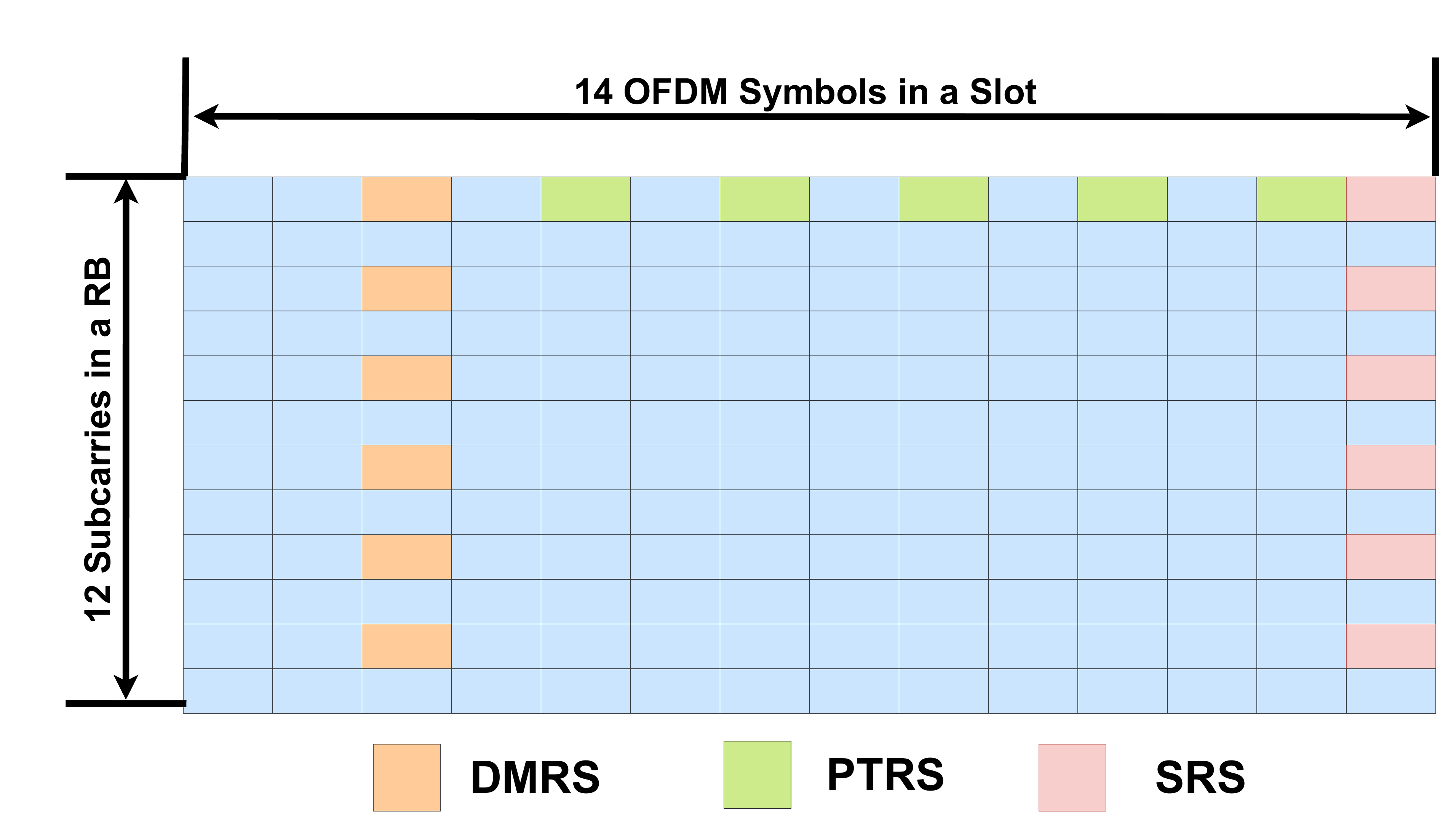}
    \caption{Uplink resource block and reference signals.}
    \label{fig:resource_block}
\end{figure}

In 5G NR, the fundamental unit of radio resource allocation is the \textit{resource block} (RB), which consists of \textbf{12 consecutive subcarriers} in the frequency domain and spans the duration of one slot in the time domain. A standard slot comprises \textbf{14 OFDM symbols}, meaning each RB contains \textbf{168 resource elements (REs)}.

To enable efficient and reliable uplink communication, 5G employs multiple \textit{Reference Signals} (RSs), including:
\begin{itemize}
    \item \textbf{DMRS}: Enables coherent demodulation of uplink data and is transmitted alongside the \textit{Physical Uplink Shared Channel} (PUSCH) or the \textit{Physical Uplink Control Channel} (PUCCH).
    \item \textbf{SRS}: Facilitates channel sounding, allowing the gNB to assess uplink channel quality and spatial characteristics, enabling beamforming and link adaptation.
    \item \textbf{Phase-Tracking Reference Signal (PTRS)}: Primarily used to mitigate phase noise and improve signal integrity in high-frequency transmissions.
\end{itemize}

Fig.~\ref{fig:resource_block} illustrates the allocation of these reference signals within a resource block.

\subsection{DMRS-SRS Relationship in Uplink CSI}

\begin{figure}[t!]
     \centering
     \begin{subfigure}{0.47\columnwidth}
         \centering
         \includegraphics[width=\textwidth,keepaspectratio]{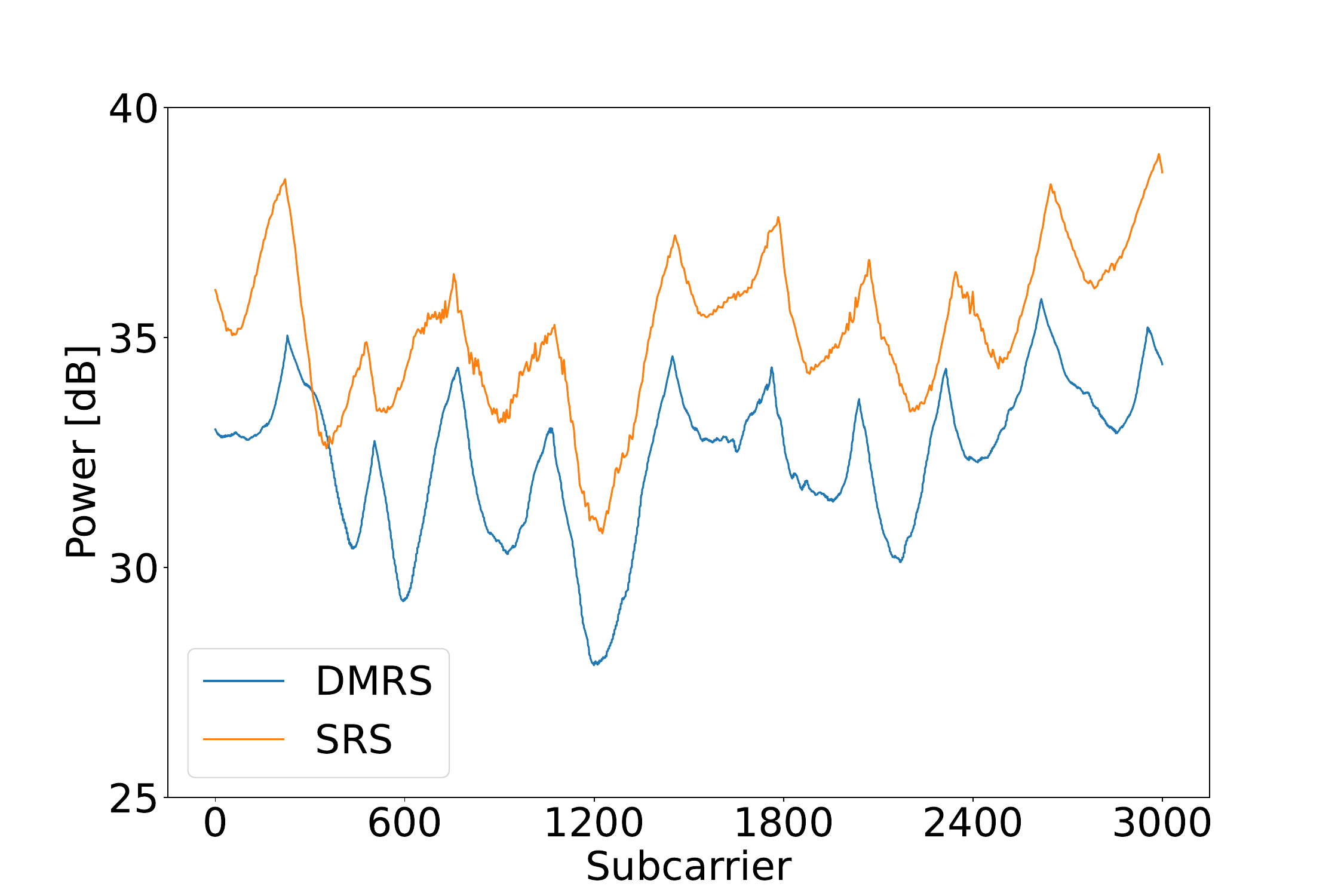}
         \caption{CSI power for DMRS and SRS in the same slot.}
         \label{fig:dmrs_srs_power}
     \end{subfigure}
     \begin{subfigure}{0.47\columnwidth}
         \centering
         \includegraphics[width=\textwidth,keepaspectratio]{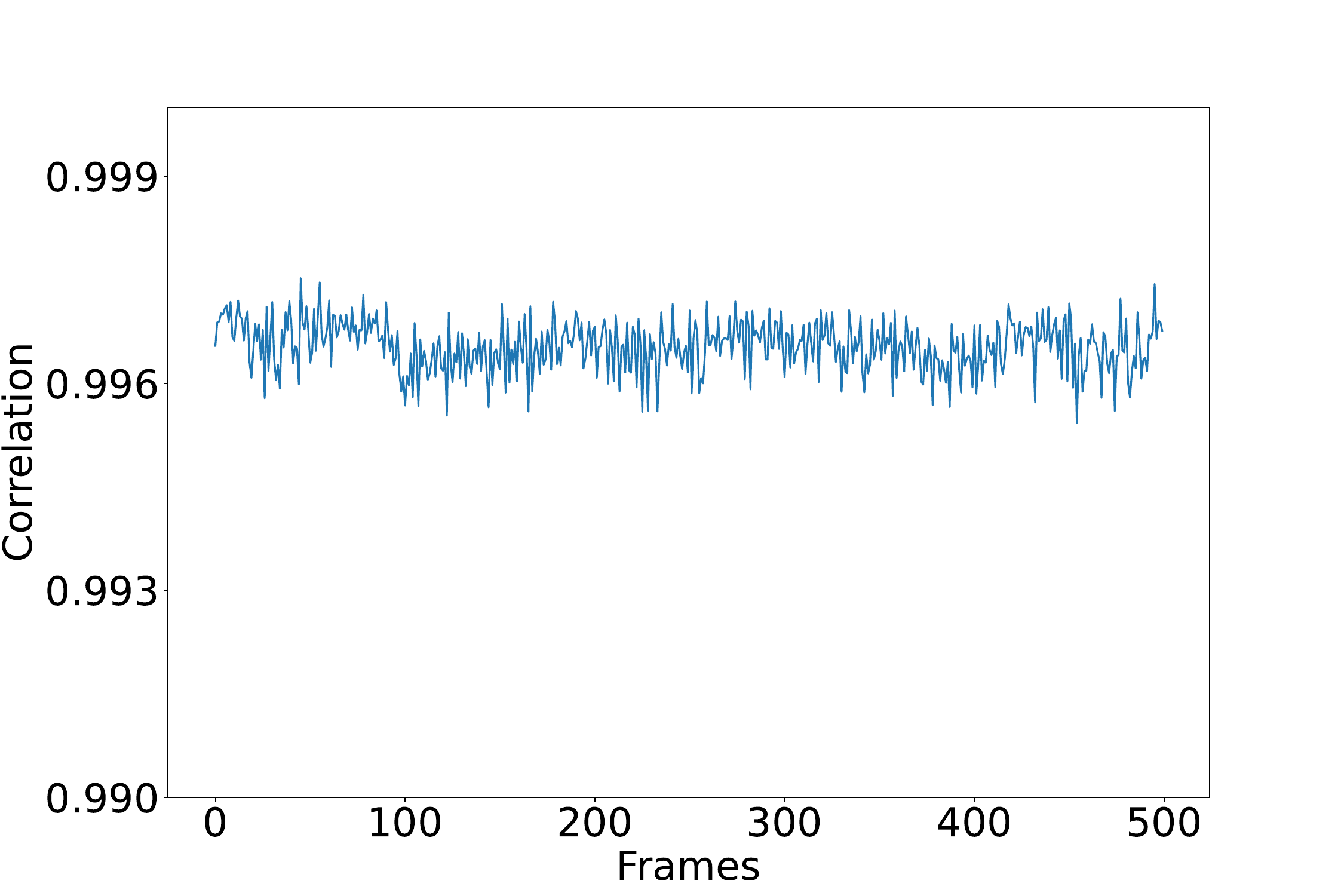}
         \caption{Correlation of DMRS and SRS power over time.}
         \label{fig:dmrs_srs_corr}
     \end{subfigure}
     \caption{Analysis of DMRS and SRS channel state information (CSI).}
\end{figure}

To demonstrate the interoperability between DMRS and SRS, we present a comparison of their CSI power levels within the same slot, shown in Fig.~\ref{fig:dmrs_srs_power}. For this experiment, the UE is configured to occupy the entire bandwidth for both data transmission (DMRS) and SRS transmission. The results reveal that while DMRS and SRS exhibit a highly similar \textit{channel gain pattern}, the primary difference lies in their respective power levels. This discrepancy stems from the distinct transmission power control mechanisms applied to PUSCH/DMRS and SRS.

Furthermore, Fig.~\ref{fig:dmrs_srs_corr} illustrates the correlation function of DMRS and SRS power across multiple frames, demonstrating the temporal consistency of their interoperability over time.

\section{Motivation and Challenges} \label{sec:motivation}

This section highlights the practical challenges of implementing ISAC in 5G  vRAN. We begin by reviewing existing 5G sensing methods, which rely on  CSI obtained from downlink or uplink reference signals~\cite{ruan2022ipos, wei20225g, gao2022toward, cha20255g}. These methods can be categorized into four types:
\begin{itemize}
    \item \textbf{Downlink Mono-static:} The base station transmits and receives signals, processing reflections for sensing.
    \item \textbf{Uplink Mono-static:} The UE handles both transmission and reception for sensing.
    \item \textbf{Downlink Bi-static:} The base station transmits while another entity (a base station via sidelink or a UE) receives.
    \item \textbf{Uplink Bi-static:} The UE transmits while a base station or another UE (via sidelink) receives.
\end{itemize}

While mono-static sensing techniques resemble traditional radar operations, they are not supported in current 5G standards and are therefore incompatible with most \textit{Commercial off-the-shelf} (COTS) 5G devices. In contrast, bi-static sensing decouples the transmitter and receiver roles, making it more feasible within 5G networks. In \textbf{uplink bi-static sensing}, the UE serves as the transmitter while the base station acts as the receiver, reducing the computational and transmission burdens on the resource-constrained UE compared to downlink sensing. Additionally, uplink bi-static sensing enables UE motion estimation (e.g., location and speed tracking)~\cite{jiang2014communicating}, which can further enhance communication performance, particularly for downlink beamforming. Therefore, this paper focuses on uplink bi-static sensing.

\subsection{Challenges in Enhancing Sensing Accuracy}

Three primary strategies exist for improving sensing accuracy:
\begin{enumerate}
    \item Increasing \textbf{signal bandwidth} enhances ranging resolution.
    \item Deploying more \textbf{antennas} improves angular resolution.
    \item Reducing the \textbf{sensing interval} enhances velocity resolution.
\end{enumerate}

However, in practical deployments, expanding the number of antennas at the gNB is challenging due to hardware constraints. Consequently, improvements must focus on optimizing \textbf{bandwidth} and \textbf{sensing intervals} in the time and frequency domains.

A recent study, \textit{ElaSe}~\cite{chen2024elase}, proposed an adaptive method to optimize SRS allocation based on user mobility, aligning sensing opportunities with UE velocity. However, this approach has three fundamental problems:

\subsubsection*{Problem 1: Scalability in SRS Allocation}

5G networks have inherently constrained SRS resources, yet they must efficiently serve a large number of UEs with varying mobility profiles. ElaSe’s adaptive SRS allocation strategy increases the sampling rate for high-mobility UEs in a linear fashion, which introduces \textbf{scalability challenges}. In dense networks with many high-speed users, resource contention can arise, making it difficult to satisfy all sensing demands within the available SRS pool. This limitation creates conflicts between demand and availability, necessitating an adaptive and scalable resource management solution.

\subsubsection*{Problem 2: Side Effects on Communication}

The sensing-oriented SRS allocation strategy of ElaSe inadvertently impacts communication performance. Since SRS is critical for accurate CSI acquisition, reducing its allocation for any given UE degrades its channel estimation quality. 
This, in turn, may lead to suboptimal \textit{Modulation and Coding Scheme} (MCS) selection, causing either overly conservative choices that underutilizes spectral efficiency or overly aggressive selections that result in high packet error rates and retransmissions.

These issues become even more pronounced in multi-user scenarios where limited SRS resources must be dynamically shared, potentially leaving some UEs with insufficient CSI updates. Consequently, while ElaSe improves sensing for high-mobility UEs, it risks \textbf{degrading communication performance} for others. A refined scheduling strategy is required to balance sensing accuracy with robust channel estimation.

\begin{figure}[t!]
    \centering
    \includegraphics[width=\columnwidth]{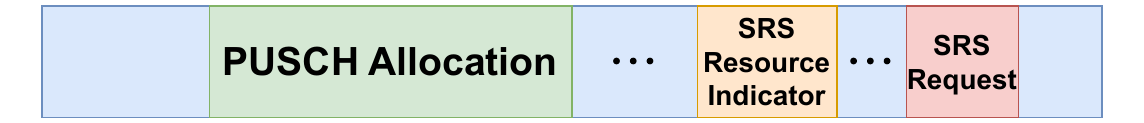}
    \caption{Illustration of DCI Format 0\textunderscore1.}
    \label{fig:dci_0_1}
\end{figure}

\subsubsection*{Problem 3: Reconfiguration Messaging Overheads}

ElaSe dynamically adjusts SRS periodicity and bandwidth to optimize sensing. However, modifying these parameters requires the gNB to send \textit{Radio Resource Control} (RRC) Reconfiguration messages to inform the UE of the new settings. This introduces signaling overhead and processing delays, resulting in latencies in the order of tens to hundreds of milliseconds~\cite{hailu2018rrc}.

Additionally, failure to acknowledge an RRC Reconfiguration message can result in link instability, reattempted signaling, or even temporary disconnection. These drawbacks make frequent updates impractical for real-time sensing.

\subsection{Our Approach: Joint DMRS-SRS Fusion}

To address these challenges, we propose a novel approach that \textbf{fuses DMRS and SRS} to enhance:
\begin{itemize}
    \item \textbf{User capacity} (scalability challenge).
    \item \textbf{Channel estimation accuracy} (communication performance challenge).
\end{itemize}

Instead of relying solely on periodic SRS transmissions, we leverage \textbf{aperiodic SRS}, an existing feature in both 4G LTE and 5G NR, which allows SRS to be triggered on demand. Specifically, we utilize \textbf{Downlink Control Information (DCI) Format 0\textunderscore1} (Fig.~\ref{fig:dci_0_1}) to trigger aperiodic SRS in conjunction with normal PUSCH allocations, which means no additional frames are needed to inform UE of its SRS allocation.
However, integrating DMRS and SRS presents additional challenges:

\subsubsection*{Challenge 1: Sporadic and Bursty PUSCH Allocations}

As discussed in \cref{uplink_ref}, DMRS and SRS play crucial roles in uplink CSI acquisition and sensing. Effective utilization of both is essential for enhancing CSI update rates. However, uncoordinated DMRS-SRS usage may lead to redundant channel estimations and inefficient use of sensing resources.

The primary challenge lies in the tight coupling between DMRS and PUSCH. Since PUSCH allocations are inherently \textbf{sporadic and bursty}, driven by base station scheduling and UE buffer status, aligning SRS with PUSCH requires careful coordination. 

A simple solution might involve modifying the PUSCH scheduling algorithm to better align with SRS. However, most RAN vendors implement proprietary, closed-source scheduling mechanisms, making direct modifications infeasible. Additionally, altering PUSCH scheduling would introduce significant complexity and could disrupt network stability.

\subsubsection*{Challenge 2: Asynchronization Between DMRS and SRS}

A major issue in fusing DMRS and SRS is the \textbf{temporal misalignment} in CSI acquisition. Since DMRS is embedded within each PUSCH transmission, it provides CSI only when uplink data is being transmitted. In contrast, aperiodic SRS is transmitted on demand and can occur independently of PUSCH allocations.
This misalignment results in inconsistent CSI estimates and potential errors in channel gain and phase estimation.

These discrepancies become particularly problematic in high-mobility environments, where channel conditions fluctuate rapidly. Existing methods for aligning CSI measurements by stitching DMRS and SRS signals sampled at different times~\cite{xiong15tonetrack} fail to account for these dynamic variations (\cref{sec:time_align}). Without proper synchronization, suboptimal scheduling decisions and inaccurate MCS selections may occur, ultimately degrading network performance.

\section{CARTS} \label{sec:method}
\label{sec:carts}

This section presents the design of \textbf{CARTS}, our proposed solution for providing CSI estimates by jointly utilizing DMRS and SRS. CARTS consists of two main components that address the challenges identified in the previous section. 

\begin{figure}[t!]
    \centering
    \hspace{0cm}\includegraphics[width=\columnwidth]{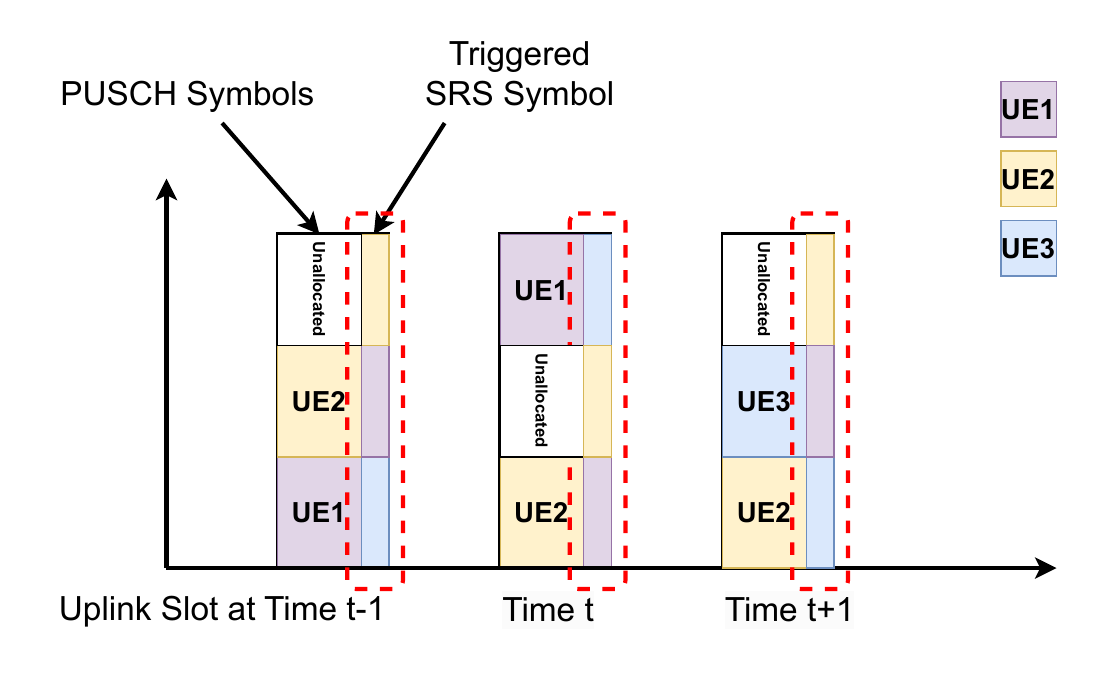}
    \caption{Illustration of Adaptive SRS Triggering.}
    \label{fig:srs_trigger}
\end{figure}

The first component (\cref{adaptive_algo}) adaptively triggers SRS to complement DMRS, reducing redundant channel sensing while ensuring that each UE receives an allocated average sensing rate. Unlike ElaSe~\cite{chen2024elase}, which allocates the entire gNB channel bandwidth (see Fig.~\ref{fig:frame_structure}) to a single UE within a time slot, CARTS partitions the bandwidth into multiple disjoint sub-bands and dynamically allocates these sub-bands to different UEs. Fig.~\ref{fig:srs_trigger} illustrates this approach, where different UEs utilize different frequency bands for SRS sounding across different time slots. 
Since DMRS and SRS channel measurements occur at different symbol times, the second component (\cref{sync_comp}) introduces a method to stitch these asynchronous CSI measurements together, ensuring accurate and consistent channel estimation.

\begin{figure}[t!]
    \centering
    \includegraphics[width=\columnwidth]{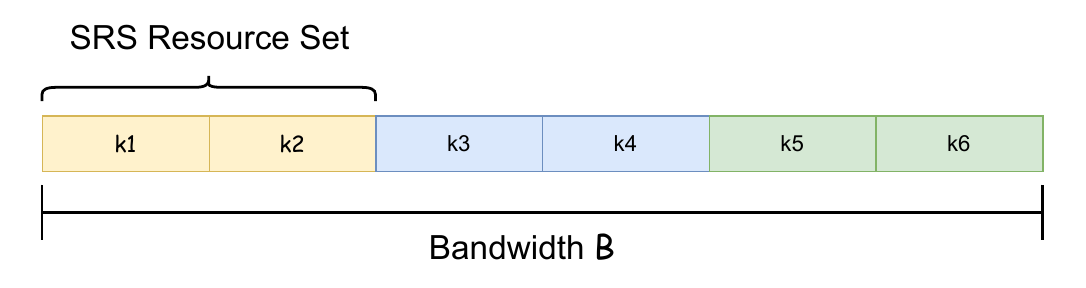}
    \caption{Pre-configured Aperiodic SRS Resources.}
    \label{fig:a_srs}
\end{figure}

\subsection{Adaptive Aperiodic SRS Triggering} \label{adaptive_algo}

\subsubsection{Pre-configured Aperiodic SRS Resources}

To facilitate coordinated sharing of the gNB channel bandwidth for SRS sounding, CARTS partitions the bandwidth into disjoint frequency sub-bands. The granularity of this partitioning is constrained by the 5G standard.

The standard mandates that aperiodic SRS bandwidth allocation follows a hierarchical structure. At the top level, the bandwidth is divided into \textbf{resource sets}, with the number of sets defined by the parameter \verb|maxNrofSRS-TriggerStates-1| in 3GPP TS 38.331~\cite{3gpp-ts-38.331}, which allows for up to three sets. Fig.~\ref{fig:a_srs} illustrates this partitioning, where each resource set is represented by a different color. Each UE can be allocated at most one resource set at a time. The gNB communicates resource set allocation to the UE using the 2-bit wide \textbf{SRS Request field} in \textbf{DCI Format 0\textunderscore1} (Fig.~\ref{fig:dci_0_1}), where the bit pattern \verb|00| indicates that no aperiodic SRS has been triggered.

At the lower level, each resource set is further divided into \textbf{resources}, determined by the parameter
\verb|maxNumberSRS|\\
\verb|-ResourcePerSet| in 3GPP TS 38.306~\cite{3gpp-ts-38.306}. This parameter can be either 1 or 2; in this paper, we assume it is set to 2 unless otherwise specified. Fig.~\ref{fig:a_srs} shows that each resource set is divided into 2 resources (e.g., the yellow resource set consists of \verb|k1| and \verb|k2|). This results in a total of six pre-configured aperiodic SRS resources (2 resources per set $\times$ 3 sets). 

The gNB can allocate both resources within a set to a single UE or assign them to two different UEs, with the constraint that each UE can have at most one resource set at a time. Consequently, a UE may be allocated between 0 and 2 resources per transmission instance, and if assigned two resources, they must belong to the same resource set. The gNB communicates these allocations using the \textbf{RS Resource Indicator (SRI) field} in DCI Format 0\textunderscore1 (Fig.~\ref{fig:dci_0_1}).

For simplicity, we assume all UEs have the same pre-configured aperiodic SRS resource configuration, as shown in Fig.~\ref{fig:a_srs}, with three resource sets and two resources per set. Additionally, we assume the gNB evenly divides its channel bandwidth among the six resources, meaning each resource occupies $\frac{1}{6}$ of the total bandwidth. \textcolor{re}{Note that our algorithm in \cref{adaptive_algo} can be extended to support heterogeneous UE configurations by having the gNB maintain a per-UE profile of its SRS capabilities (i.e., srs\_bw\_configs), which can be queried during the resource allocation in Algorithm 1. This would involve a more complex data structure, but it does not fundamentally change the priority-based allocation logic.}

\subsubsection{Sensing QoS: Target Channel Estimation Rate} \label{sec:tgt_rate}

As discussed in \cref{sec:motivation}, combining DMRS (within PUSCH) and aperiodic SRS enhances channel sensing. However, relying solely on DMRS can result in extended gaps in CSI updates, particularly for UEs with sporadic or bursty traffic. 

To ensure sufficiently frequent channel updates, we introduce the \textbf{target channel estimation rate}, denoted as \(\text{tgt\_rate}[u]\) for each UE~\(u\). This predefined rate is set based on the UE’s mobility and traffic characteristics:
\begin{itemize}
    \item High-mobility or bursty-traffic UEs: \(\text{tgt\_rate}[u] = 200\) estimates/second.
    \item Stationary or low-traffic UEs: \(\text{tgt\_rate}[u] = 50\) estimates/second.
\end{itemize}

Although the target rate is predefined, achieving it is not always guaranteed due to resource constraints. Moreover, CARTS dynamically utilizes \textbf{all} available SRS resources (\cref{sec:a_srs_alloc}), meaning UEs may receive higher update rates when resource contention is low. Fine-tuning the target rate based on specific UE communication and sensing requirements, and optimizing SRS allocation for power efficiency are left for future work.

\subsubsection{DMRS-Aware Aperiodic SRS Allocation Algorithm} \label{sec:a_srs_alloc}

The goal of the proposed aperiodic SRS allocation algorithm is to efficiently distribute SRS resources among UEs to meet their respective target channel estimation rates across the channel bandwidth while ensuring that:
\begin{enumerate}
    \item No two UEs use the same SRS resource simultaneously.
    \item Each UE is assigned only one resource set at a time.
\end{enumerate}

\textbf{Algorithm Notation:}
\begin{itemize}
    \item \( U \) - Set of all connected UEs.
    \item \( R \) - Set of all Resource Blocks (RBs) (Fig.~\ref{fig:frame_structure}).
    \item \( K = \{1, 2, ..., 6\} \) - Set of six SRS resources with the configuration in Fig.~\ref{fig:a_srs}.
    \item \( \text{D}_{k, k'} \) - Indicator variable; 1 if resources \( k \) and \( k' \) belong to different resource sets, 0 otherwise.
    \item \( \text{pusch\_alloc}[u, r] \) - Indicates if UE \( u \) is using RB \( r \) in the latest PUSCH allocation.
    \item \( \text{last\_est\_time}[u, r] \) - Stores the last time UE \( u \) performed a channel estimation in RB \( r \), and is a real matrix of dimension $|U| \times |R|$.
\end{itemize}

The algorithm prioritizes UEs based on how delayed their last channel estimation is relative to their target rate. A greedy selection process iteratively assigns the most urgent UE-resource pair, ensuring fairness while maximizing CSI update efficiency.
The detailed algorithm is shown in \cref{alg:srs_allocation_custom}.
It begins by updating $\text{last\_est\_time}$ with the latest PUSCH allocation $\text{pusch\_alloc}$ in Line \ref{alg:line:update_last_est_time_with_pusch} where $t$ is the slot time. After that, in Line 
\ref{alg:update_ch_est_pri}, the quantity $t - \text{last\_est\_time}[u, r]$ tells how long ago UE $u$ made a channel measurement in RB $r$. If this quantity is larger than $\frac{1}{\text{tgt\_rate}[u]}$, then it means UE $u$ is behind the expected channel estimation rate for RB $r$ and vice versa. Thus, a large positive number in $\text{ch\_est\_pri[u,r]}$ (Line \ref{alg:update_ch_est_pri}) means the channel estimation is behind schedule and should be given the priority. In Line \ref{alg:comp_urgency_matrix}, the matrix element $\text{srs\_bw\_configs}[r,k]$ is 1 if the RB $r$ is in resource $k$ and 0 otherwise, i.e. the $|R|$-by-6 matrix indicates whether a RB is in a SRS resource. So, a large value in $\text{urgency\_matrix}[u, k]$ in Line \ref{alg:comp_urgency_matrix} means it is more urgent for UE $u$ to use SRS resource $k$ for channel estimation. Given the computed urgency, Line \ref{alg:find_most_urgent} uses greedy selection to select the pair $(\hat{u},\hat{k})$ with the maximum urgency. Since SRS resource $\hat{k}$ has been allocated to a selected UE, the other UEs cannot use it (Line \ref{alg:exclude_other_UE}) and the selected UE cannot use other resource sets (Line \ref{alg:exclude_other_resource_sets}).

\algnewcommand\algorithmicvariable{\textbf{variable: }}
\algnewcommand\VARIABLE{\State \algorithmicvariable}

\begin{algorithm}[ht]
\caption{Aperiodic SRS Triggering Allocation}
\label{alg:srs_allocation_custom}
\begin{algorithmic}[1]
\Function{generate\_srs\_alloc}{}


\ForAll{$(u, r) \text{ such that pusch\_alloc[u, r] is 1}$}
    \State $\text{last\_est\_time}[u,r] \gets \text{t}$ \label{alg:line:update_last_est_time_with_pusch}
    \EndFor \Comment{Update last\_est\_time from PUSCH}

\ForAll{$u \in U$ \text{ and } $r \in R$}
    \State $\text{ch\_est\_pri}[u,r] \! \gets \! t  - \text{last\_est\_time}[u,r]
            - \tfrac{1}{\text{tgt\_rate}[u]}$ \label{alg:update_ch_est_pri}
\EndFor \Comment{Compute channel-est. priority}

\State $\text{urgency\_matrix} \gets \text{ch\_est\_pri} \times \text{srs\_bw\_configs}$ \label{alg:comp_urgency_matrix}

\State $\text{SRS\_resource\_alloc} \gets [\,]$
\While{$\max(\text{urgency\_matrix})> -\infty$}
  \State $(\hat{u}, \hat{k}) \gets \arg\max(\text{value\_matrix})$ \label{alg:find_most_urgent}

  \State \text{append} $(\hat{u}, \hat{k})$ \text{ to }\text{SRS\_resource\_alloc}

  \State $\text{urgency\_matrix}[u, \hat{k}] \gets -\infty \quad \forall u \in U$
  \label{alg:exclude_other_UE}
  
 \State $\text{urgency\_matrix}[\hat{u}, \hat{k}] \gets -\infty \quad \forall k \text{ that } D_{k,\hat{k}}=1$ \label{alg:exclude_other_resource_sets}

\EndWhile \Comment{Greedy selection: pick max}



\State Update $\text{last\_est\_time}$ with $\text{SRS\_resource\_alloc}$
\State \Return $\text{SRS\_resource\_alloc}$, $\text{last\_est\_time}$
\EndFunction
\end{algorithmic}
\end{algorithm}

\subsection{CSI Synchronization and Compensation} \label{sync_comp}

Recent studies on channel stitching in high-bandwidth wireless systems emphasize the importance of merging partial channel estimates to construct a unified and accurate representation of the wireless channel. A notable example is \textit{ToneTrack}~\cite{xiong15tonetrack}, originally designed for overlapping Wi-Fi channels. ToneTrack treats adjacent channels as a special case of overlapping channels, first correcting the intrinsic phase slope across each channel’s subcarriers, then aligning the phase between the last subcarrier of one channel and the first subcarrier of the next, effectively stitching them together. 

While Wi-Fi-based channel stitching techniques provide valuable insights, directly applying these methods to 5G NR presents several unforeseen challenges. A key distinction lies in the mobility assumptions: Wi-Fi stitching techniques typically assume a relatively static transmitter and receiver, focusing primarily on phase stitching. In contrast, 5G networks are inherently designed for mobility. In the uplink scenario, the UE moves within the coverage area while the base station remains stationary, dynamically altering multipath characteristics such as:
\begin{itemize}
    \item \textbf{Spatial Directions:} Changes in the relative position of the moving UE affect signal arrival angles (\cref{sec:spatial_smooth}).
    \item \textbf{Path Delays:} Variations in UE movement impact delay profiles (\cref{sec:time_align}).
    \item \textbf{Fading Profiles:} Rapid fluctuations in wireless channels alter frequency-dependent channel gains (\cref{sec:freq_calib}).
\end{itemize}

To accurately stitch the channel, we introduce a comprehensive CSI synchronization and compensation framework that:
\begin{enumerate}
    \item Smooths CSI across multiple antennas to maintain spatial consistency.
    \item Compensates for discrepancies in power levels (as shown in Fig.~\ref{fig:dmrs_srs_power}).
    \item Corrects the phase misalignment in CSI across adjacent DMRS and SRS sub-bands.
\end{enumerate}

We define each CSI estimate obtained from DMRS or SRS as a 3-tuple \(\big({H}_b, \mathcal{N}_b, t_b\big)\), where:
\begin{itemize}
    \item \( {H}_b \in \mathbb{C}^{M \times N_b} \) represents the complex CSI matrix for sub-band \( b \), where \( M \) is the number of receiving antennas, and \( N_b = |\mathcal{N}_b| \) is the number of subcarriers in sub-band $b$.
        \item \( \mathcal{N}_b \) denotes the set of subcarriers used in DMRS or SRS for that measurement.
    \item \( t_b \) is the time at which the CSI estimate was obtained.
\end{itemize}

Channel stitching begins once all subcarriers have been measured at least once (i.e., when \(\bigcup_{b} \mathcal{N}_b\) covers the full bandwidth) and the latest measurement originates from an SRS triggering event. We designate this SRS sub-band as the \textbf{reference sub-band} \( b_{\rm ref} \). 
\textcolor{re}{Note that both SRS and DMRS contain unused subcarriers along the frequency domain within a resource block, as illustrated in Fig.~\ref{fig:resource_block}. In such cases, channel estimates for these subcarriers can be readily obtained through interpolation.}

\subsubsection{Spatial-Domain Smoothing} \label{sec:spatial_smooth}

In mobile scenarios, the relative position and orientation of the UE (as well as scatterers in the environment) change over time, causing variations in the spatial signature of the received signal across different antennas. To ensure spatial coherence, we leverage the spatial correlation across antennas by constructing a weighted covariance matrix of CSI vectors from all sub-bands and extracting principal components based on measurement time and bandwidth.

Let \( H_b(n) \in \mathbb{C}^{M \times 1} \) denote the CSI vector for subcarrier \( n \) in sub-band \( b \), and let \( t_{\rm ref} = \max_{b} \{t_{b}\} \) be the most recent measurement time. We define a decay parameter \( \alpha > 0 \) and compute the time-weighted coefficient \( w_b \) as:
\begin{equation}
    w_b = \exp(-\alpha (t_{\rm ref} - t_b)).
\end{equation}
The weighted spatial covariance matrix is then given by:
\begin{equation} \label{eq:weighted_cov}
\mathbf{R}
\;=\;
\frac{1}{\sum_{b}\sum_{n} w_{b}}
\sum_{b}
\sum_{n}
w_{b} H_b(n) H_b(n)^{H},
\end{equation}
where \((\cdot)^H\) denotes the Hermitian transpose.

Performing eigen-decomposition on \(\mathbf{R}\) yields eigenvalues \(\{\lambda_i\}\) and corresponding eigenvectors \(\{\mathbf{u}_i\}\). The principal eigenvector \(\mathbf{u}_1\) (associated with the largest eigenvalue \(\lambda_1\)) represents the dominant spatial mode of the channel. We then project each sub-band’s CSI vector onto \(\mathbf{u}_1\) and update $H_b$ using the following phase correction:
\begin{equation}
    H_b \leftarrow e^{-j\,\angle(\mathbf{u}_1^H H_b)} H_b.
\end{equation}
This ensures all sub-bands share a common spatial reference, mitigating phase inconsistencies across sub-bands.

\subsubsection{Time-Domain Alignment} \label{sec:time_align}

Time-domain alignment corrects temporal offsets between sub-bands. We first obtain the \textit{channel impulse response} (CIR) by applying the inverse discrete Fourier transform (IDFT) to \( H_b \). We denote the delay tap with index $k$, and the peak delay \( \tau_b \) for sub-band $b$ is identified as:

\begin{equation}
    \tau_b 
    \;=\; \arg\max_{-\frac{N_b}{2} \,\le\, k \,<\,\frac{N_b}{2}} 
    \bigl\lvert h_b[k] \bigr\rvert.
    \label{eq:peak_estimation}
\end{equation}


Since different sub-bands may be measured at different times, their peak delays \( \tau_b \) can be misaligned. We define the \textbf{global reference delay} as:
\begin{equation}
    \tau_{\rm ref} = \max_{b} \tau_b.
\end{equation}
To align all CIRs, we apply a delay shift:
\begin{equation}
    h_b'[k] = h_b[k - (\tau_{\rm ref} - \tau_b)].
\end{equation}

\begin{figure}
     \centering
     \begin{subfigure}{\columnwidth}
         \hspace{0cm}
         \includegraphics[width=\textwidth,keepaspectratio]{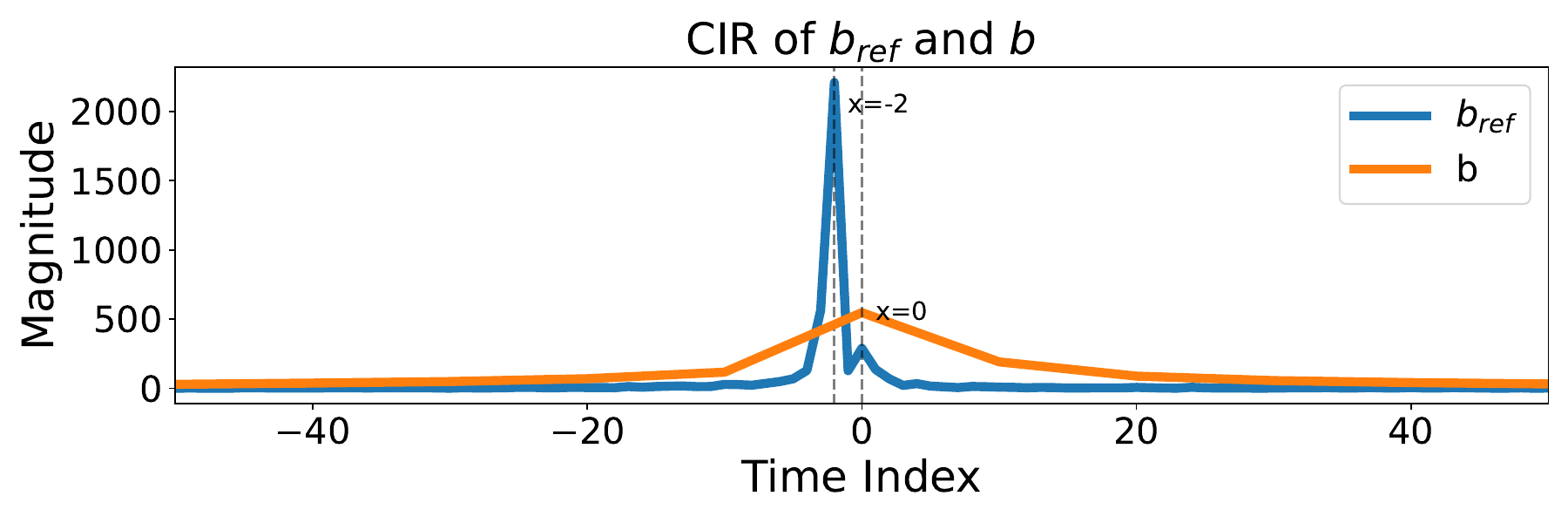}
         \caption{}
         \label{fig:cir_res}
     \end{subfigure}
     \begin{subfigure}{0.32\columnwidth}
         \centering
         \includegraphics[width=\textwidth,keepaspectratio]{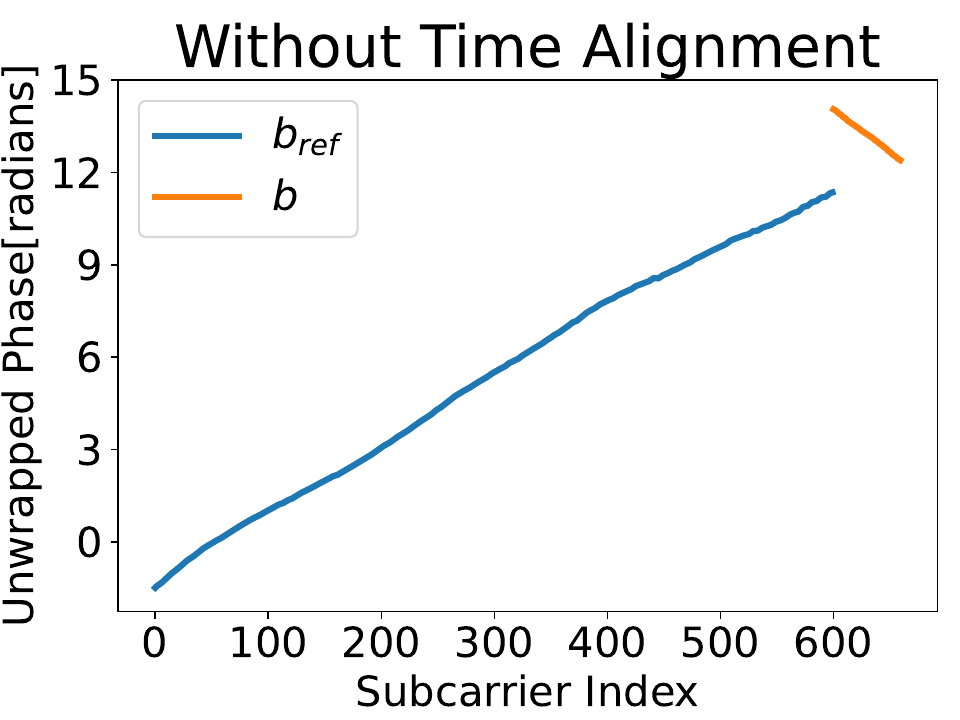}
         \caption{}
         \label{fig:no_align_res}
     \end{subfigure}
     \hfill
     \begin{subfigure}{0.32\columnwidth}
         \centering
         \includegraphics[width=\textwidth,keepaspectratio]{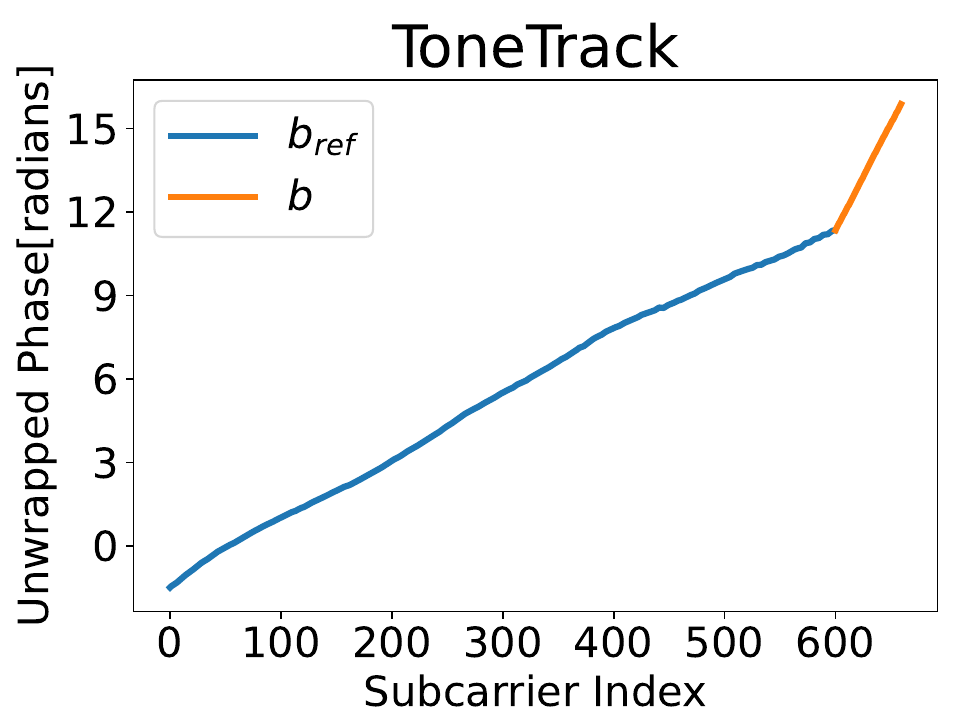}
         \caption{}
         \label{fig:tk_align_res}
     \end{subfigure}
     \hfill
     \begin{subfigure}{0.32\columnwidth}
         \centering
         \includegraphics[width=\textwidth,keepaspectratio]{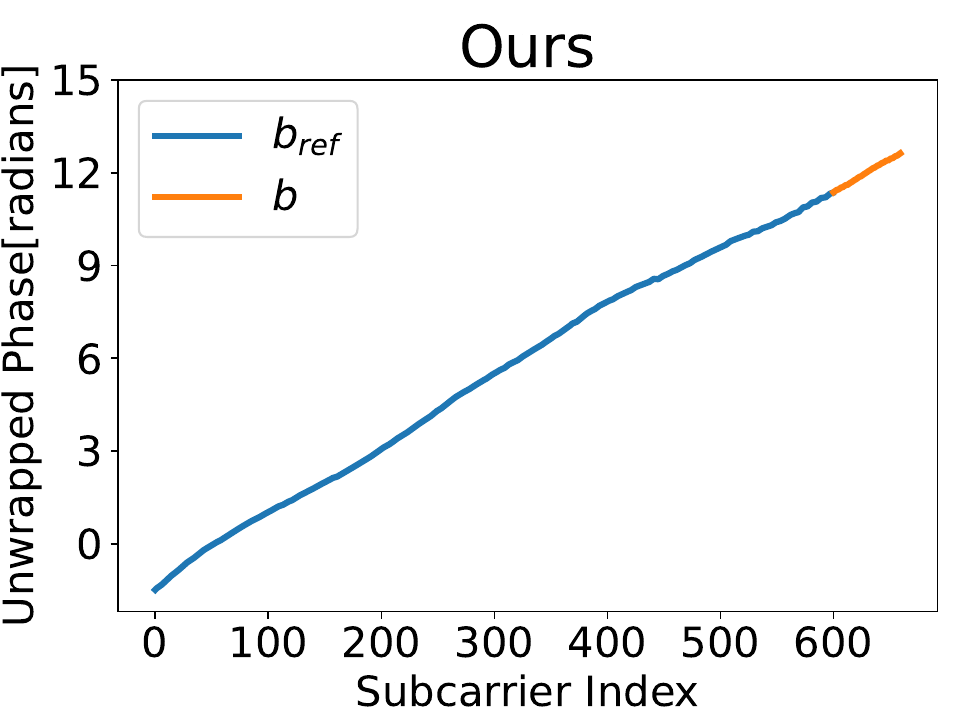}
         \caption{}
         \label{fig:ours_align_res}
     \end{subfigure}
     \caption{(a) CIR of \( b_{\rm ref} \) and \( b \), (b) Phases of \( b_{\rm ref} \) and \( b \) without time alignment, (c) Phase of aligned sub-bands using ToneTrack, (d) Phase of aligned sub-bands using our slope-based method.}
     \vspace{-5mm}
\end{figure}

A key challenge in sub-band measurements arises when each sub-band has a \textsl{narrow} bandwidth, and only a limited number of RBs are typically active in the uplink direction.  
Figure~\ref{fig:cir_res} presents example CIRs from real CSI traces for two sub-bands. The reference band spans 50 RBs (600 subcarriers), whereas the other sub-band \( b \) occupies only 5 RBs. As illustrated in Fig.~\ref{fig:no_align_res}, these sub-bands exhibit different phase slopes prior to alignment.


Although methods such as \textit{ToneTrack}~\cite{xiong15tonetrack} and \textcolor{re}{HiSAC~\cite{pegoraro2024hisac}} attempt to align the CIR peak of sub-band~$b$ with that of a reference sub-band~$b_{\mathrm{ref}}$ (see Fig.~\ref{fig:tk_align_res}), this alignment is often imprecise due to the wide delay spread associated with narrow sub-band bandwidths. Moreover, even small errors in peak-based alignment can accumulate when stitching multiple sub-bands, leading to significant inaccuracies in timing advance estimation (from the CIR) and angle estimation (via the steering vector).

To address this issue, we introduce a new strategy that avoids direct CIR peak shifting. Instead, we eliminate the linear phase slope from each sub-band while retaining the slope \( \alpha_{b_{\mathrm{ref}}} \) and phase intercept \( \phi_{b_{\mathrm{ref}},0} \) from the reference band. Let \( H_b(n) \) represent the measured channel coefficient for sub-band \( b \) at subcarrier index \( n \), and let \( n_0 \) denote the first subcarrier index of that sub-band. We model the phase of \( H_b(n) \) as:  
\begin{equation}
    \arg H_b(n) \approx \phi_{b,0} + \alpha_b (n - n_0),
    \label{eq:phase_model}
\end{equation}  
where \( \alpha_b \) is the estimated phase slope, and \( \phi_{b,0} \) is the intercept. By fitting a line to the phase values across \( n \in \mathcal{N}_b \), we estimate \( \alpha_b \) and \( \phi_{b,0} \). We then remove this slope via:  
\begin{equation}
    \hat{H}_b(n) = H_b(n)e^{-j\alpha_b(n - n_0)}.
    \label{eq:slope_correction}
\end{equation} 
\textcolor{re}{The reason behind this is: ideally, perfect time alignment would result in both sub-bands sharing the same phase slope, and under Line-of-Sight (LOS) conditions, the unwrapped phase will be almost linear.
This operation also ensures each sub-band retains small phase curvatures containing multi-path information while aligning all sub-bands.} In the next step, we stitch \( \hat{H}_b \) to \( \hat{H}_{b_{\rm ref}} \).

\subsubsection{Frequency-Domain Compensation} \label{sec:freq_calib}

To align sub-band \( b \) with the reference sub-band \( b_{\text{ref}} \), we introduce a complex scaling factor \( \gamma_b \):
\begin{equation}
    \gamma_b = \beta_b e^{j\phi_b},
    \label{eq:gamma_def}
\end{equation}
where \( \beta_b \) and \( \phi_b \) represent the amplitude and phase offsets, respectively.

A straightforward approach to estimate \( \beta_b \) and \( \phi_b \) is to leverage overlapping subcarriers, i.e., \( n \in \mathcal{N}_b \cap \mathcal{N}_{b_{\mathrm{ref}}} \):
\begin{align}
    \beta_b 
    &= \sqrt{\frac{\sum\limits_{n \in \mathcal{N}_b \cap \mathcal{N}_{b_{\mathrm{ref}}}} |\hat{H}_{b_{\mathrm{ref}}}(n)|^2}
       {\sum\limits_{n \in \mathcal{N}_b \cap \mathcal{N}_{b_{\mathrm{ref}}}} |\hat{H}_b(n)|^2}},
    \label{eq:beta_estimate}
    \\
    \phi_b 
    &= \arg \sum\limits_{n \in \mathcal{N}_b \cap \mathcal{N}_{b_{\mathrm{ref}}}} 
       \hat{H}_{b_{\mathrm{ref}}}(n) \odot \hat{H}_b^{*}(n),
    \label{eq:phi_estimate}
\end{align}
where \( \odot \) denotes element-wise multiplication.

If no overlapping subcarriers exist (\( \mathcal{N}_b \cap \mathcal{N}_{b_{\mathrm{ref}}} = \varnothing \)), we estimate \( \beta_b \) and \( \phi_b \) using an adjacent sub-band \( b \). Since we assume that all subcarriers have been estimated at least once, an adjacent sub-band is always available.  

Let \( b_{\mathrm{ref}} \) be the reference band with boundary indices  
\( (n_{\mathrm{start}}^{(b_{\mathrm{ref}})}, n_{\mathrm{end}}^{(b_{\mathrm{ref}})}) \)  
and let \( b \) be the adjacent sub-band with  
\( (n_{\mathrm{start}}^{(b)}, n_{\mathrm{end}}^{(b)}) \).  
We define the boundary subcarrier pair \( (n_{\mathrm{boundary}}^{(b_{\mathrm{ref}})}, n_{\mathrm{boundary}}^{(b)}) \) based on their relative positions:
\[
(n_{\mathrm{boundary}}^{(b_{\mathrm{ref}})}, n_{\mathrm{boundary}}^{(b)}) =
\begin{cases}
(n_{\mathrm{end}}^{(b_{\mathrm{ref}})}, n_{\mathrm{start}}^{(b)}), & \text{if } b_{\mathrm{ref}} < b,\\
(n_{\mathrm{start}}^{(b_{\mathrm{ref}})}, n_{\mathrm{end}}^{(b)}), & \text{otherwise}.
\end{cases}
\]

After removing each channel’s internal phase slope, we estimate the amplitude and phase offsets as:
\begin{align}
    \beta_b 
    &= \frac{|\hat{H}_{b_{\mathrm{ref}}}(n_{\mathrm{boundary}}^{(b_{\mathrm{ref}})})|}
         {|\hat{H}_b(n_{\mathrm{boundary}}^{(b)})|},        
    \label{eq:unified_beta}
    \\[4pt]
    \phi_b
    &= \arg\hat{H}_{b_{\mathrm{ref}}}(n_{\mathrm{boundary}}^{(b_{\mathrm{ref}})}) 
    - \arg\hat{H}_b(n_{\mathrm{boundary}}^{(b)}).
    \label{eq:unified_phi}
\end{align}

Once \( \beta_b \) and \( \phi_b \) are determined, we align sub-band \( b \) to the reference \( b_{\mathrm{ref}} \) using:
\begin{equation}
    \hat{H}_b'(n) = \gamma_b \hat{H}_b(n) = \beta_b e^{j\phi_b} \hat{H}_b(n).
    \label{eq:calibrated_CSI}
\end{equation}

Here, \( \hat{H}_b'(n) \) is now both amplitude- and phase-aligned with the reference sub-band \( b_{\text{ref}} \). For overlapping subcarriers, we retain the measurement with the most recent timestamp. Specifically, the stitched reference channel estimate  
\( \hat{H}_{\mathrm{ref}}'(n) \) is defined over the full subcarrier set \( \mathcal{N}_b \cup \mathcal{N}_{b_{\mathrm{ref}}} \) as:
\begin{equation}
\label{eq:keep_most_recent}
\hat{H}_{\mathrm{ref}}'(n)
=
\begin{cases}
\hat{H}_b'(n), 
    & \text{if } n \in \mathcal{N}_b \setminus \mathcal{N}_{b_{\mathrm{ref}}},\\[6pt]
\hat{H}_{b_{\mathrm{ref}}}(n), 
    & \text{if } n \in \mathcal{N}_{b_{\mathrm{ref}}} \setminus \mathcal{N}_b,\\[6pt]
\hat{H}_b'(n), 
    & \text{if } n \in \mathcal{N}_b \cap \mathcal{N}_{b_{\mathrm{ref}}} \text{ and } t_b > t_{\mathrm{ref}},\\[6pt]
\hat{H}_{b_{\mathrm{ref}}}(n),
    & \text{otherwise}.
\end{cases}
\end{equation}

In the overlapped region \( \mathcal{N}_b \cap \mathcal{N}_{b_{\mathrm{ref}}} \), the most recent measurement is always used, leveraging the timestamps \( t_b \) and \( t_{\mathrm{ref}} \). This ensures that for each subcarrier \( n \), the final stitched estimate \( \hat{H}_{\mathrm{final}}(n) \) preserves the most up-to-date CSI information.

\subsubsection{Iterative Compensation for Full-Band Stitching}

The compensation process described above aligns and stitches a single sub-band \( b \) with the reference \( b_{\text{ref}} \). To construct a unified channel estimate across the entire bandwidth, we iteratively apply this process. After each successful stitching operation, the newly formed composite channel \( \hat{H}_{\mathrm{ref}}'(n) \) becomes the updated reference for subsequent iterations.

To minimize error propagation, we prioritize adjacent sub-bands and proceed outward from the initial reference \( b_{\text{ref}} \) in both frequency directions. This iterative process continues until all available sub-bands are incorporated into a single, coherent channel estimate that preserves the temporal and spectral dynamics of the most recent measurements.

The final stitched channel estimate provides a seamless representation of the full-band CSI, effectively mitigating amplitude and phase mismatches while maintaining spectral consistency across the entire frequency range.

\subsubsection{Interpolation for Uniform Sampling}

To ensure uniform channel estimation intervals, we apply interpolation. Given a target rate \( Y \), we use spline interpolation to resample stitched estimates:
\begin{equation}
\hat{H}_s''(n, k) = \mathrm{spline}\Bigl(\bigl\{t_j, \hat{H}_s'(n, t_j)\bigr\}_{j=0}^{N-1}\Bigr)\Bigl(t_0 + \frac{k}{Y}\Bigr).
\end{equation}

This ensures consistent channel tracking across different UEs and time instances. Fig.~\ref{fig:ours_align_res} illustrates the result of the proposed CSI synchronization and compensation for example sub-bands, demonstrating significantly higher accuracy compared to \textit{ToneTrack} (Fig.~\ref{fig:tk_align_res}).

\begin{figure}[ht]
    \centering
    \begin{subfigure}{0.45\columnwidth}
        \centering
        \includegraphics[width=\textwidth,keepaspectratio]{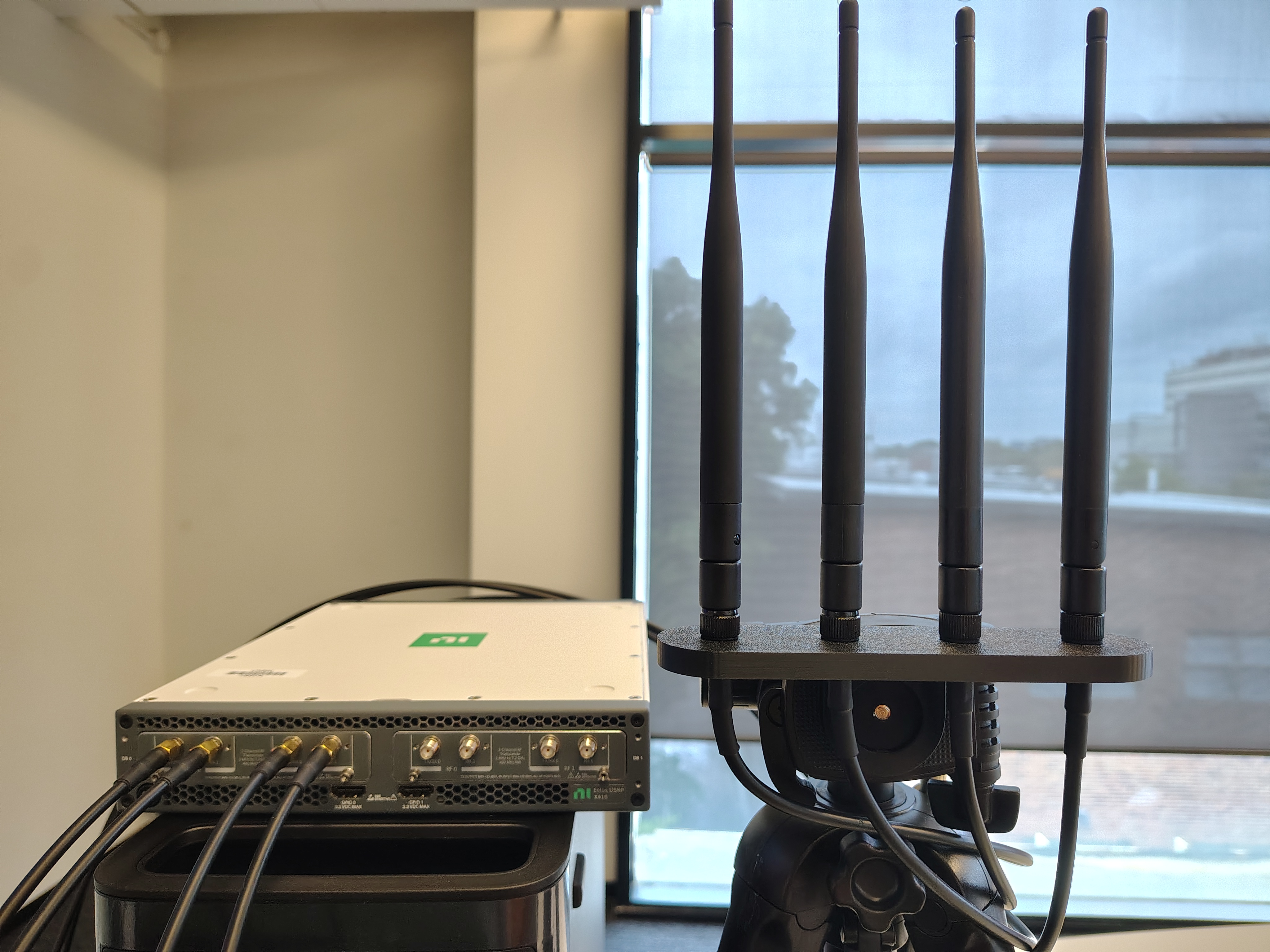}
        \caption{}
    \end{subfigure}
    \hspace{0.5cm}
    \begin{subfigure}{0.45\columnwidth}
        \centering
        \includegraphics[width=\textwidth,keepaspectratio]{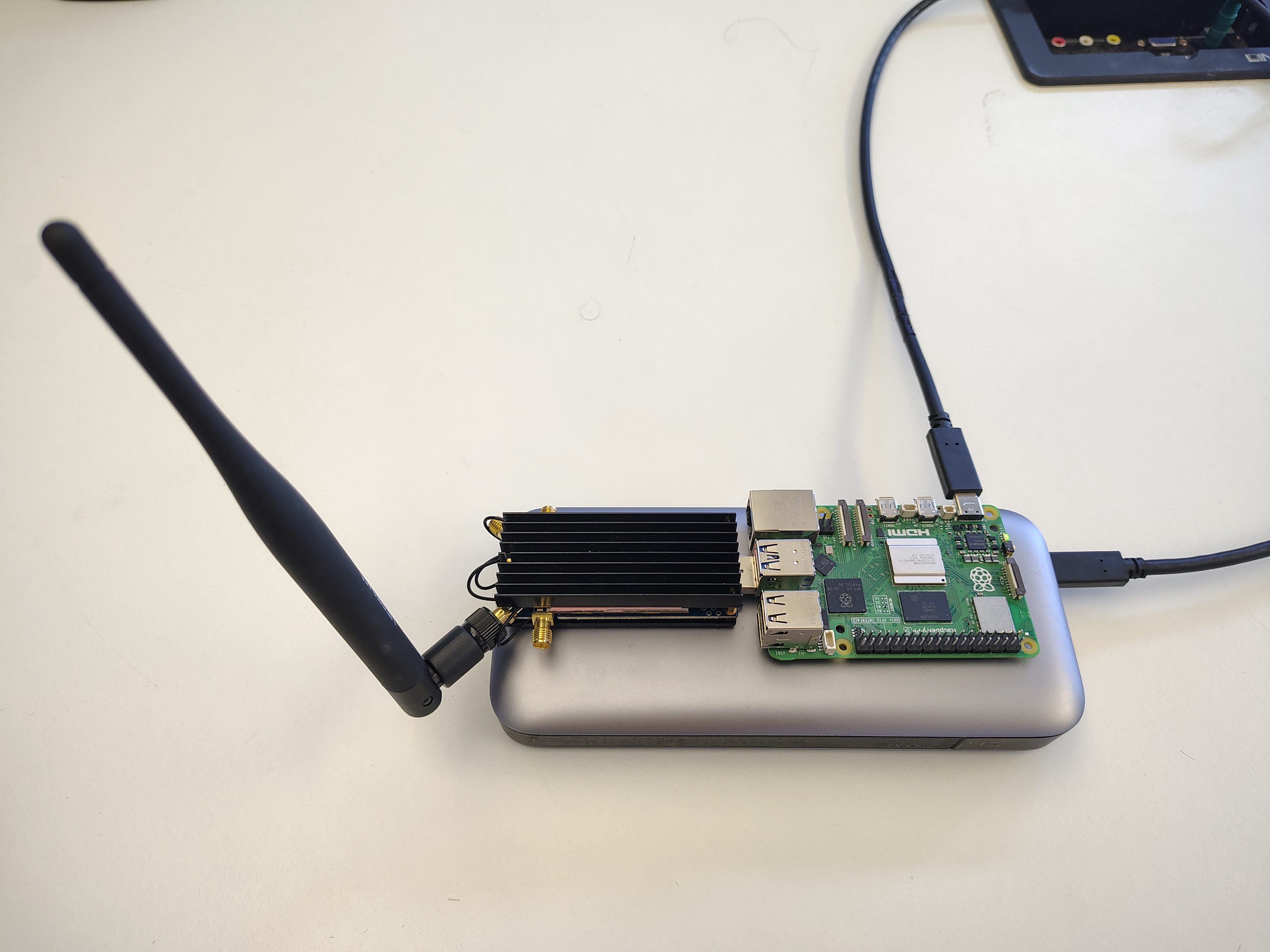}
        \caption{}
    \end{subfigure}
     \vspace{0mm}
    \caption{(a) Base station and antenna array, (b) Quectel RM530N module.}
    \label{fig:testbed}
     \vspace{0mm}
\end{figure}

\section{EVALUATION}

\begin{figure*}[t!]
    \centering
    \includegraphics[width=\linewidth]{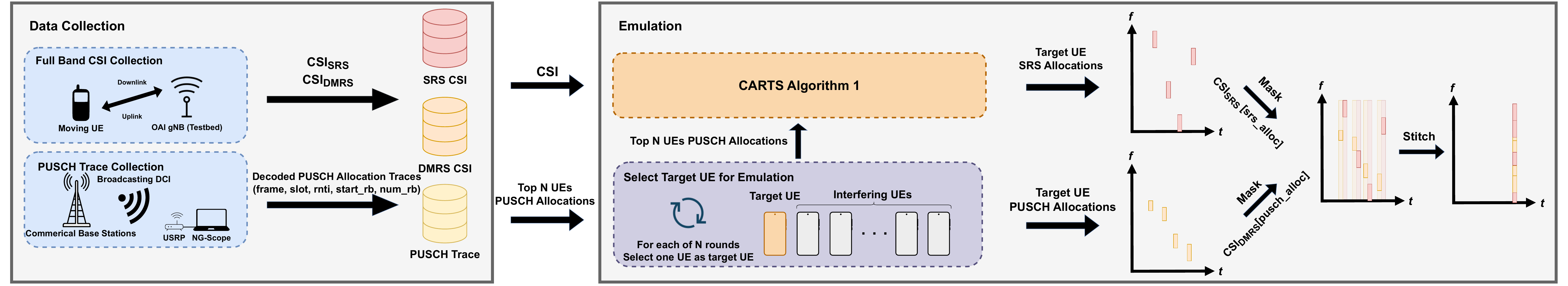}
    \caption{\textcolor{re}{Overview of the trace-driven emulation framework. We first collect full-band CSI from an OAI testbed and PUSCH allocation traces from commercial networks. In the emulation, these traces drive an N-user scenario where the CARTS algorithm uses PUSCH information to trigger SRS allocations. The corresponding DMRS and SRS CSI snippets are then retrieved and stitched to reconstruct the channel for a target UE.}}
    \label{fig:eval_process}
\end{figure*}

To evaluate our proposed scheme under realistic conditions, we developed a testbed comprising a USRP X410 (equipped with two transmit and two receive antennas) running the open-source 5G stack \textit{OpenAirInterface} (OAI) and an RM530N Quectel module acting as the UE (Fig.~\ref{fig:testbed}). The base station operates in the 5G FR1 band at 4.0 GHz (Band 77), utilizing a 100 MHz bandwidth with a subcarrier spacing of 30 kHz, yielding a total of 273 RBs. Since the current OAI release does not support aperiodic SRS triggering, we primarily used this setup to collect CSI for offline emulation.


\textcolor{rectc}{Fig.~\ref{fig:eval_process} illustrates the components of our emulation framework. The left side of the figure represents the data collection phase, where: (i) we use the testbed to collect CSI measurements, and (ii) we employ a sniffing tool to capture real-world PUSCH allocations from commercial mobile networks. On the right side of Fig.~\ref{fig:eval_process}, the collected data is used to generate CSI measurements for the target user. We provide further details on these components below.}




\textcolor{rectc}{To ensure the realism of our emulation, we aim to incorporate realistic DMRS allocations.} 
\textcolor{re}{To achieve this, we utilize cellular sniffer tools to extract DCI from nearby base stations, thereby obtaining PUSCH allocation traces. Unfortunately, existing open-source 5G sniffers support only either Standalone (SA) \cite{wan2024nr} or Frequency Division Duplexing (FDD) \cite{5gsniffing2023} deployments. In contrast, all detectable base stations in our experimental environment operate under 5G Non-Standalone (NSA) with Time Division Duplexing (TDD), which these sniffers do not support. As a result, we opted to use the 4G sniffer \textit{NG-Scope}~\cite{xie2022ng} to collect PUSCH traces from nearby 4G base stations.}

Since 4G uplink transmissions are limited to 100 resource blocks (RBs), we scale the recorded PUSCH allocations by a factor of 2.72 and round the result to match the 272 RBs available in our emulation, while also ensuring that the final RB count is a multiple of 4. 
\textcolor{re}{Each collected PUSCH trace is formatted as $(\textit{frame}, \textit{slot}, \textit{rnti}, \textit{start\_rb}, \textit{num\_rb})$, where \textit{rnti} denotes the unique identifier of each UE.}

\begin{figure}[th]
    \centering
    \includegraphics[width=\columnwidth]{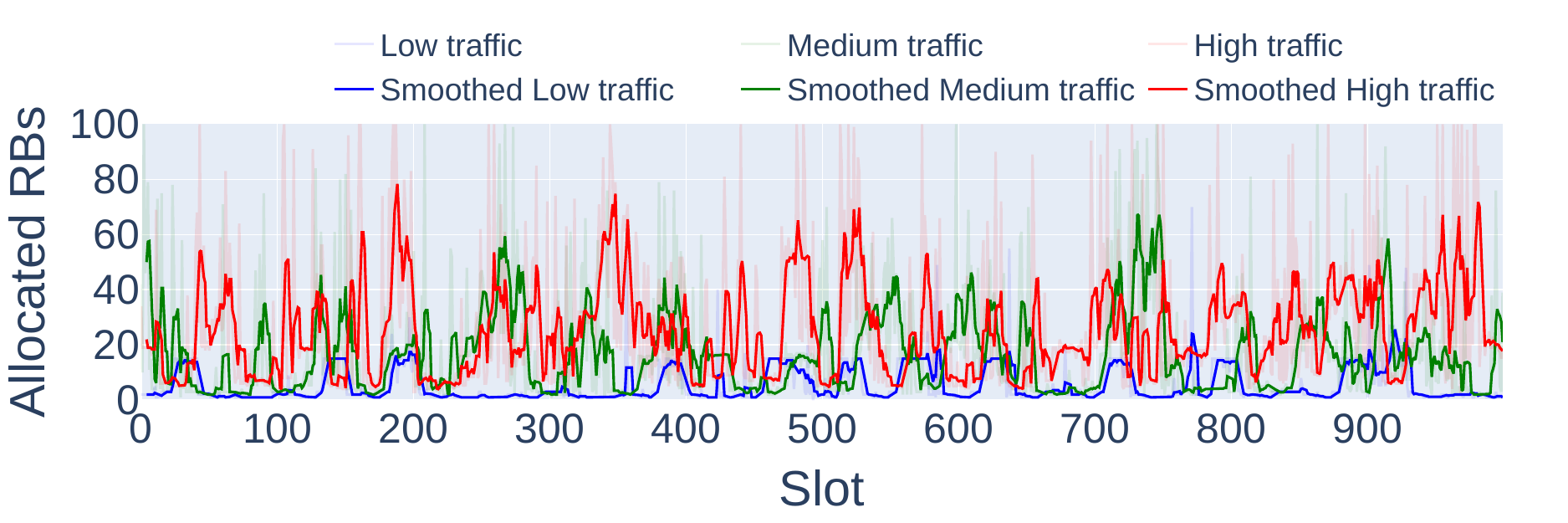}
     \vspace{-3mm}
    \caption{Uplink traffic from real PUSCH traces.}
    \label{fig:traffic}
     \vspace{-3mm}

\end{figure}

We gathered traces from three distinct environments: a \textbf{residential area} (low traffic), a \textbf{university campus} (medium traffic), and a \textbf{city center} (high traffic). Fig.~\ref{fig:traffic} displays the smoothed traffic patterns over 1,000 slots, where the number of active UEs ranges from 100 to 300. These traces serve as the input to simulate the SRS triggering behavior of our algorithm (\cref{adaptive_algo}).

To further validate our approach, we collected CSI data using our testbed in two distinct indoor environments:
\begin{itemize}
    \item A \textbf{cluttered office} environment, representing typical indoor office scenarios.
    \item A \textbf{large open-floor room}, simulating industrial factory conditions.
\end{itemize}

Additionally, we positioned the base station and UE in separate office rooms to evaluate communication performance under \textbf{Non-Line-of-Sight (NLOS)} conditions. \textcolor{re}{Since we opt for a trace-driven approach,} \textbf{full-band CSI was collected by assigning all RBs to a single UE}, enabling the retrieval of complete-band CSI from both DMRS and SRS. During CSI collection, the UE was held and moved by a subject at approximately 3 km/h to reflect typical indoor movement scenarios~\cite{3gpp-tr-38.913}. In each environment, the UE followed three different trajectories to capture the variations in channel conditions due to movement. 


Since the number of active UEs influences SRS triggering (\cref{adaptive_algo}), we selected the top \( N \) UEs with the highest PUSCH allocations as active UEs during emulation. Unless otherwise specified, we set \( N = 10 \), following the guidelines for indoor scenarios outlined in 3GPP TR 38.913~\cite{3gpp-tr-38.913}. 
\textcolor{rectc}{When we use $N$ users in an emulation, we run $N$ rounds. The $N$ different UEs take turns to become a \textsl{target UE} over these $N$ rounds. In each round, there is a target UE while the remaining $N$-1 UEs act as interfering traffic, consuming SRS resources based on our algorithm and their assigned PUSCH traffic patterns from the traces. In each round, we apply CARTS to stitch the CSIs from DMRS and SRS of the target UE and evaluate the quality of the stitched CSIs. The PUSCH allocations for the $N$ users (which are obtained from the sniffed traces) are used as the input to Algorithm \ref{alg:srs_allocation_custom} to obtain the SRS allocations for these $N$ users (Fig.~\ref{fig:eval_process}). For the target UE, we need its CSIs at its PUSCH and SRS allocations. Let ${\rm CSI}_{\rm DMRS}$ be the measured full-band DMRS CSI from the testbed. If the PUSCH allocation for the target UE at time $t$ includes sub-carrier $s$, then we define a mask (or the indicator function): ${\rm pusch\_alloc}[t, s] = 1$, otherwise 0. The DMRS CSIs for the target UE at time $t$ at sub-carrier $s$ is ${\rm CSI}_{\rm DMRS}[t, s]$ if ${\rm pusch\_alloc}[t, s]$ is 1.  Similarly, we extract the SRS CSI for the target UE from the full-band SRS CSI measurements using its SRS allocation as a mask (Fig.~\ref{fig:eval_process}). 
}


\textcolor{re}{
 We report the average performance of the target UE across all N rounds. For the interfering UEs: half were assigned as \textbf{moving} and the rest as \textbf{stationary}, where \textbf{moving} and \textbf{stationary UEs} are assumed to have a target channel estimation rate of 200 and 50 estimations/second, respectively.}

\subsection{Communication Performance}

To evaluate the communication performance of our approach, we employ two key metrics that capture the fidelity of CSI estimation. High CSI fidelity directly improves MCS selection, optimizing spectral efficiency while ensuring reliable communication. Accurate channel estimation allows the base station to select the highest feasible MCS while maintaining acceptable error rates, thereby maximizing throughput.

To quantify CSI fidelity, we assess:

\begin{enumerate}
    \item \textbf{Normalized Mean Squared Error (NMSE):} Measures the discrepancy between the amplitude of stitched CSI from our scheme and the ground truth CSI (obtained from full-band SRS), reflecting overall channel estimation accuracy. NMSE is defined as:
     $  \text{NMSE} = \frac{\|H_{\text{true}} - H_{\text{estimated}}\|^2}{\|H_{\text{true}}\|^2}$,
    where \( H_{\text{true}} \) represents the actual channel matrix, and \( H_{\text{estimated}} \) denotes our reconstructed channel estimate.

    \item \textbf{CIR Peak Position Error:} Assesses errors in CIR peak positions, which directly impact timing advance (TA) calculations and synchronization accuracy in 5G networks. Accurate TA ensures that uplink transmissions from different UEs arrive synchronously at the base station, preventing inter-symbol interference. We evaluate this metric by comparing the strongest CIR peak positions derived from ground truth CSI and reconstructed (stitched) CSI, effectively validating our time alignment method (\cref{sec:time_align}).
\end{enumerate}

\begin{figure}[t!]
\centering
    \begin{subfigure}[b]{0.3\textwidth}
        \centering
        \includegraphics[width=\linewidth]{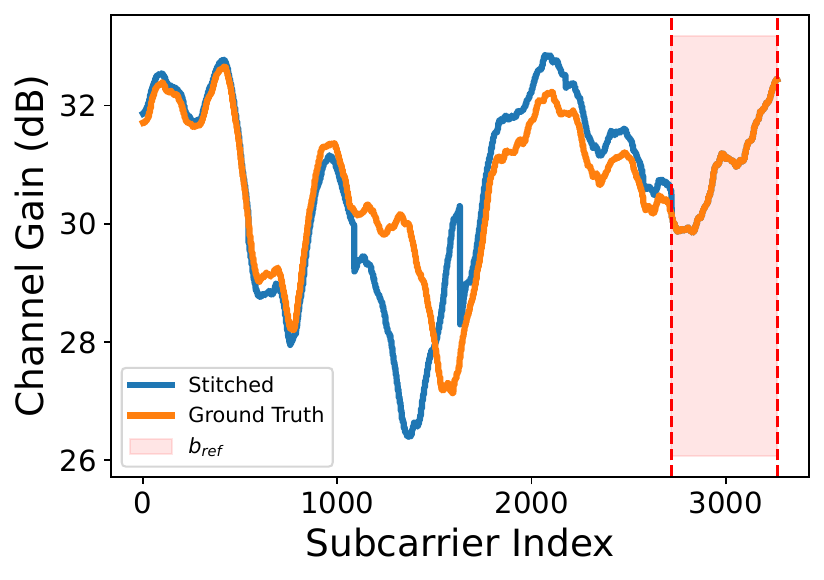}
        \caption{Example result of stitched CSI gain}
        \label{fig:chest_vis}
    \end{subfigure}
    \hfill
    \begin{subfigure}[b]{0.3\textwidth}
        \centering
        \includegraphics[width=\linewidth]{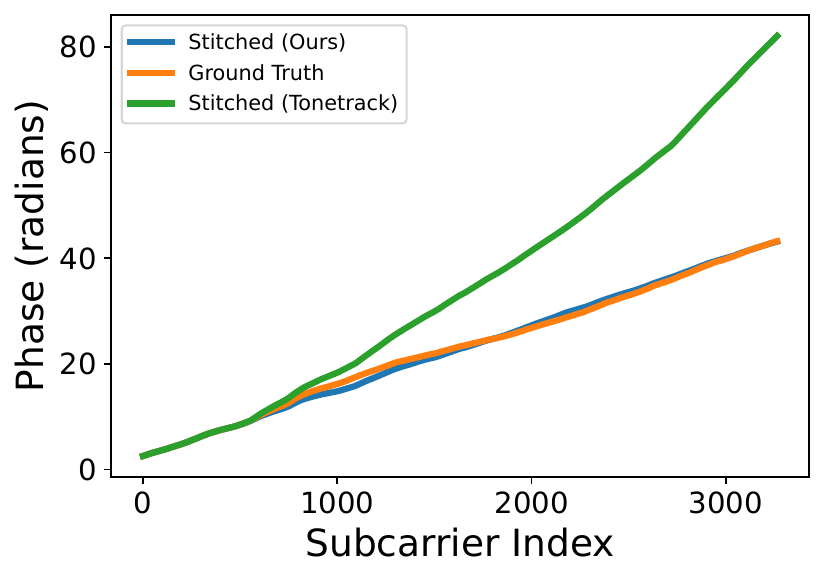}
        \caption{Example result of stitched CSI unwrapped phase}
        \label{fig:phase_vis}
    \end{subfigure}
    \caption{\textcolor{re}{Visualization of the CARTS stitching results vs. Ground Truth CSI vs. ToneTrack baseline}}
    \label{fig:qual_res}

\end{figure}

These metrics collectively evaluate how our adaptive SRS triggering scheme impacts communication reliability while preserving sensing performance. 
\textcolor{re}{In addition, we provide the qualitative results in ~Fig.~\ref{fig:qual_res}. Compared with the ground truth, we can see that the stitched CSI shares a similar fading profile with the ground truth, and the phase differences are negligible. However, ToneTrack shows significant phase drifts due to the phase error accumulation discussed in \cref{sec:time_align}.}

\begin{figure*}[t!]
    \centering
    \begin{subfigure}[b]{0.33\textwidth}
        \centering
        \includegraphics[width=\linewidth]{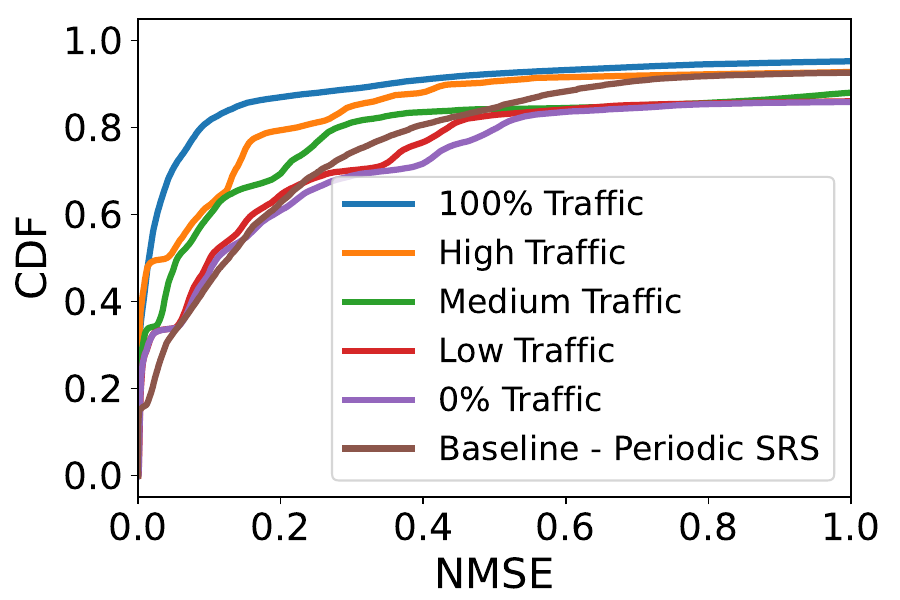}
        \caption{CDF of NMSE for varying traffic levels}
        \label{fig:nmse_results}
    \end{subfigure}
    \hfill
    \begin{subfigure}[b]{0.33\textwidth}
        \centering
        \includegraphics[width=\linewidth]{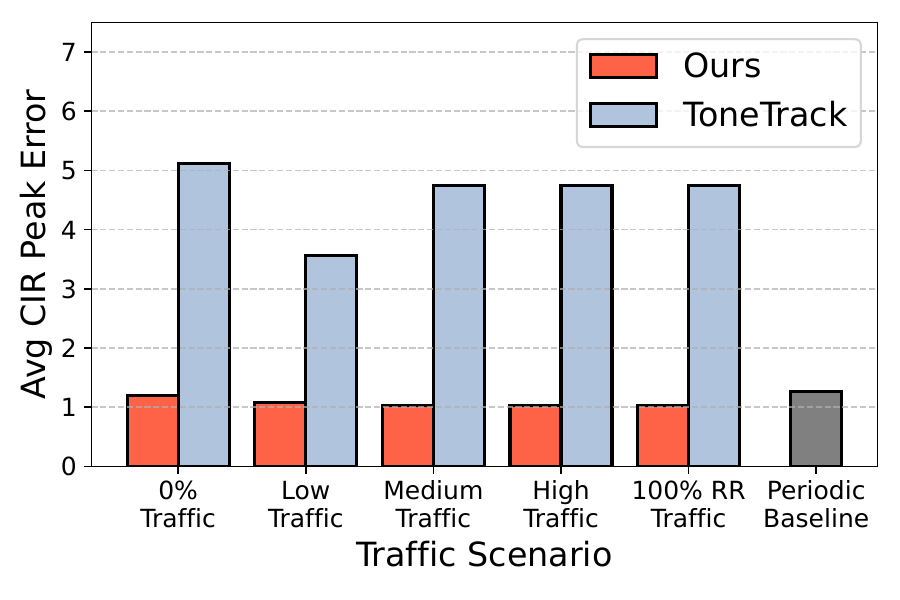}
        \caption{Average CIR peak position error}
        \label{fig:cir_peak_error}
    \end{subfigure}
    \hfill
    \begin{subfigure}[b]{0.33\textwidth}
        \centering
        \includegraphics[width=\linewidth]{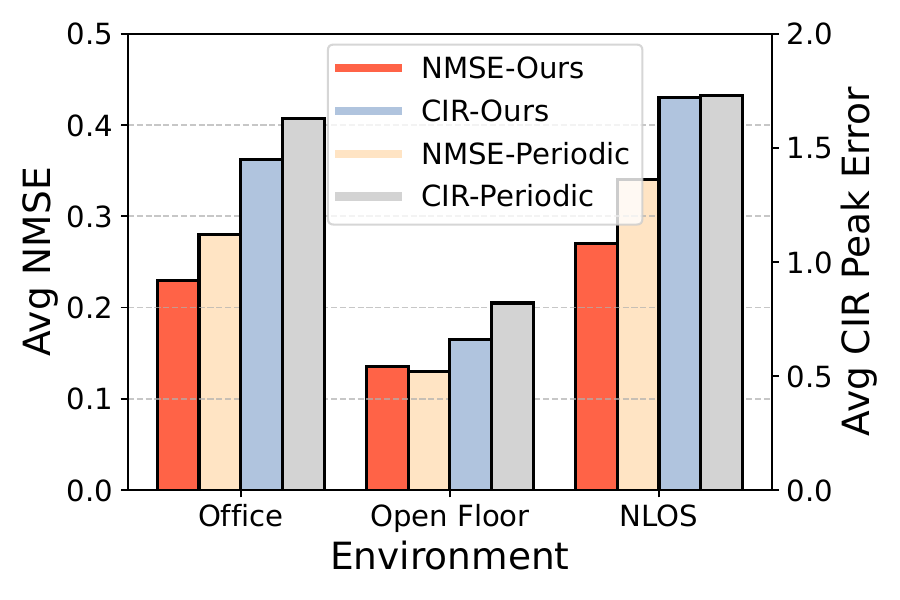}
        \caption{Performance in different environments}
        \label{fig:nmse_cir_env}
    \end{subfigure}
    \hfill
    \begin{subfigure}[b]{0.33\textwidth}
        \centering
        \includegraphics[width=\linewidth]{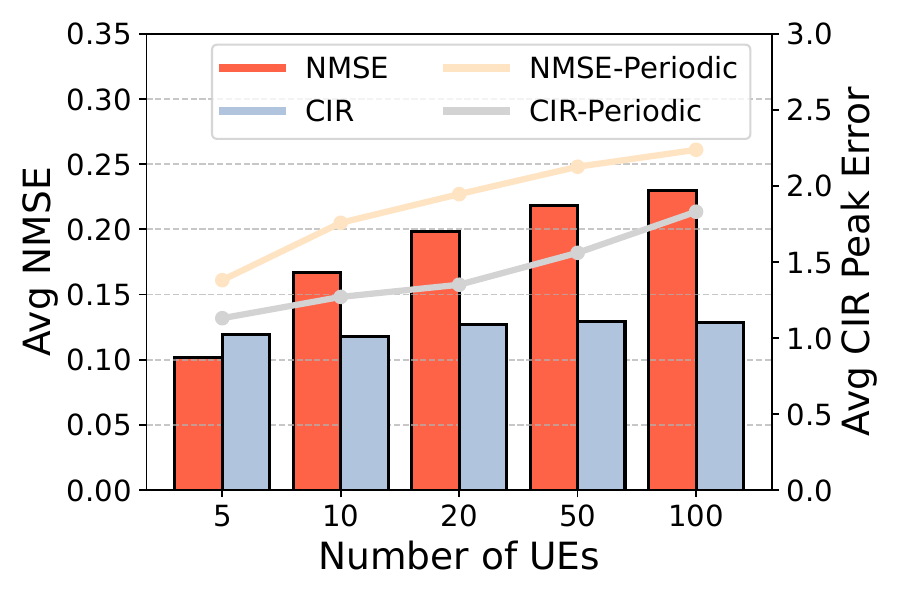}
        \caption{Performance with varying UE numbers}
        \label{fig:nmse_cir_ue}
    \end{subfigure}
    \hspace{5mm}
    \begin{subfigure}[b]{0.33\textwidth}
        \centering
        \includegraphics[width=\linewidth]{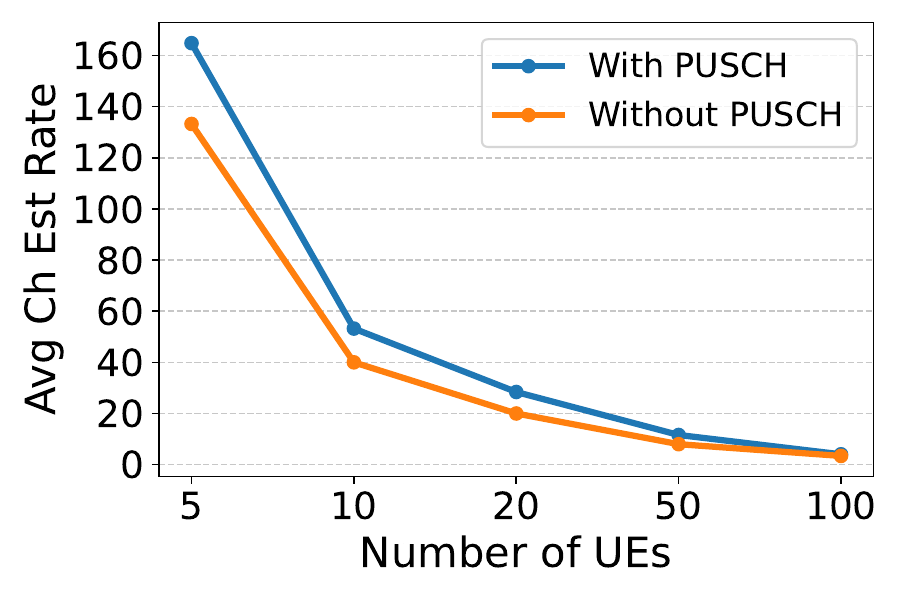}
        \caption{Average channel estimation rate}
        \label{fig:ch_est_rate}
    \end{subfigure}

    \vspace{-2mm}
    \caption{Communication performance evaluation under various conditions.}
      \vspace{-2mm}
    \label{fig:comm_performance}
\end{figure*}

\subsubsection{Normalized Mean Squared Error}

In addition to real traffic traces, we use two extreme traffic levels: \textbf{0\% traffic} (no RBs allocated in PUSCH), and \textbf{100\% traffic} (all RBs evenly allocated in PUSCH to UEs in each slot).
Fig.~\ref{fig:nmse_results} illustrates NMSE distributions under different traffic conditions. Higher traffic loads result in lower NMSE values due to increased sensing opportunities, leading to more frequent and accurate channel estimates.

Our results demonstrate that the adaptive scheme maintains median NMSE values well below 0.25 (i.e., 3 dB), even in zero- and low-traffic scenarios. This error is significantly smaller than the SNR-CQI mapping step size shown in Figure A.4 of TR 36.942~\cite{ETSI_TR_136942_2024}, indicating that channel stitching has minimal impact on MCS selection. 

Fig.~\ref{fig:nmse_results} also compares our method to a baseline using periodic SRS-only channel estimation, where periodicity is determined based on the number of active UEs, with SRS resources allocated in a round-robin manner.
Our adaptive scheme demonstrates notable performance gains in medium- and high-traffic scenarios typical of indoor office and factory environments. Under low-traffic conditions, the channel stitching process introduces minor errors, slightly increasing NMSE compared to periodic SRS.

\subsubsection{CIR Peak Position Error}

Fig.~\ref{fig:cir_peak_error} presents the average CIR peak position error under different conditions. The results indicate an average error of approximately 1 sample, with 90\% of estimates having an error below 2 samples. At a 30 kHz subcarrier spacing and 100 MHz bandwidth, this corresponds to a timing error of approximately 16.3 ns—well within the acceptable range (TA steps of 16 samples) for maintaining 5G synchronization.

Performance degradation is minimal across different traffic scenarios. As shown in Fig.~\ref{fig:cir_peak_error}, our approach significantly outperforms \textit{ToneTrack}, which exhibits a substantially higher peak position error. The baseline periodic SRS-only shows slightly worse results than ours since our SRS triggering and stitching is more frequent than the periodic method so we can get more up-to-date channel delay estimations.

\subsubsection{Performance Across Different Environments}

We further evaluate our adaptive SRS triggering scheme in different environments under high-traffic. As shown in Fig.~\ref{fig:nmse_cir_env}:
\begin{itemize}
    \item In \textbf{(LOS) conditions}, NMSE in the cluttered office (0.23) is higher than in the open-floor room (0.14), likely due to increased multipath effects.
    \item In \textbf{NLOS conditions}, NMSE rises slightly to 0.27 in spit of more complex fading processes, demonstrating that our adaptive scheme effectively estimates the channel across diverse environments.  
\end{itemize}
The baseline method works well under an open-floor environment, however, the error increases significantly in the office and NLOS conditions.
Similarly, CIR peak position errors follow a similar trend -- higher in cluttered office spaces and NLOS conditions -- but remain well within acceptable limits for 5G synchronization. The baseline periodic SRS shows higher CIR peak error under all circumstances.

\subsubsection{Performance with Varying Number of UEs}

We also assess performance under varying numbers of active UEs (\( N \)) in high-traffic conditions. Fig.~\ref{fig:nmse_cir_ue} shows that NMSE increases from 0.102 to 0.235 as \( N \) increases from 5 to 100. This is expected, as higher \( N \) values introduce greater resource contention, reducing sensing opportunities per UE. This trend is further supported by Fig.~\ref{fig:ch_est_rate}, which shows the corresponding decrease in average channel estimation rate. Despite this, the CIR peak position error remains stable across different \( N \) values, indicating that our adaptive scheme effectively maintains synchronization accuracy even with a large number of active UEs.

\textcolor{re}{We also report the periodic SRS method results under different numbers of UEs. We compare CARTS against the periodic SRS baseline, which allocates full-band SRS to $N$ UEs in a round-robin fashion. This baseline represents an \emph{ElaSe}\cite{chen2024elase}-like strategy, where the effective sensing rate for a given UE decreases as $N$ increases. Note that we cannot directly compare our method with \emph{Elase} since its required sensing rate is unachievable in our multi-UE setting due to the sensing resource contention. Our results in Fig.~\ref{fig:nmse_cir_ue} compare CARTS to this strategy under varying UE numbers. We can clearly see that our method always achieves lower errors in both metrics compared with periodic-based SRS. Our method consistently yields lower errors across both metrics compared to the periodic-based SRS approach. Remarkably, it maintains comparable performance even when supporting more number of UEs. For instance, our method achieves an NMSE of 0.167 with 10 UEs, whereas the periodic-based SRS records an NMSE of 0.161 with only 5 UEs.}

\subsection{Sensing Performance}

\begin{figure*}[t!]
    \centering
    \begin{subfigure}[b]{0.24\textwidth}
        \centering
        \includegraphics[width=\linewidth]{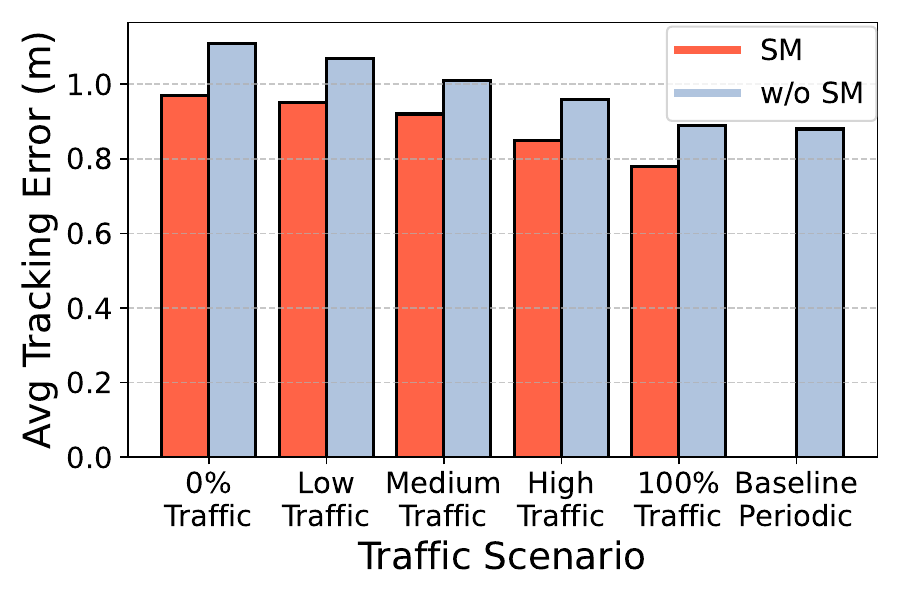}
        \caption{Average UE localization error under different traffics}
        \label{fig:loc_error}
    \end{subfigure}
    \hfill
    \begin{subfigure}[b]{0.24\textwidth}
        \centering
        \includegraphics[width=\linewidth]{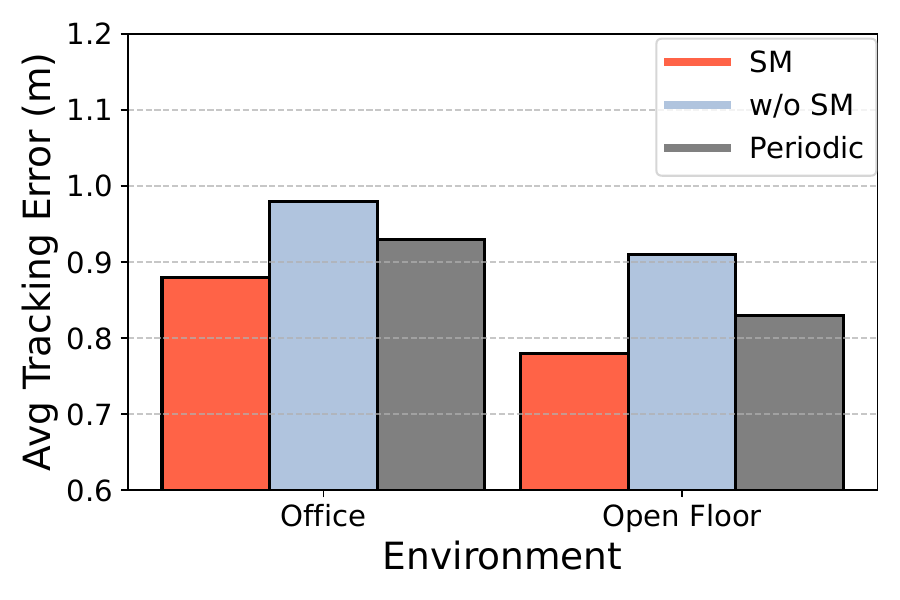}
        \caption{Performance in different environments}
        \label{fig:loc_error_env}
    \end{subfigure}
    \hfill
    \begin{subfigure}[b]{0.24\textwidth}
        \centering
        \includegraphics[width=\linewidth]{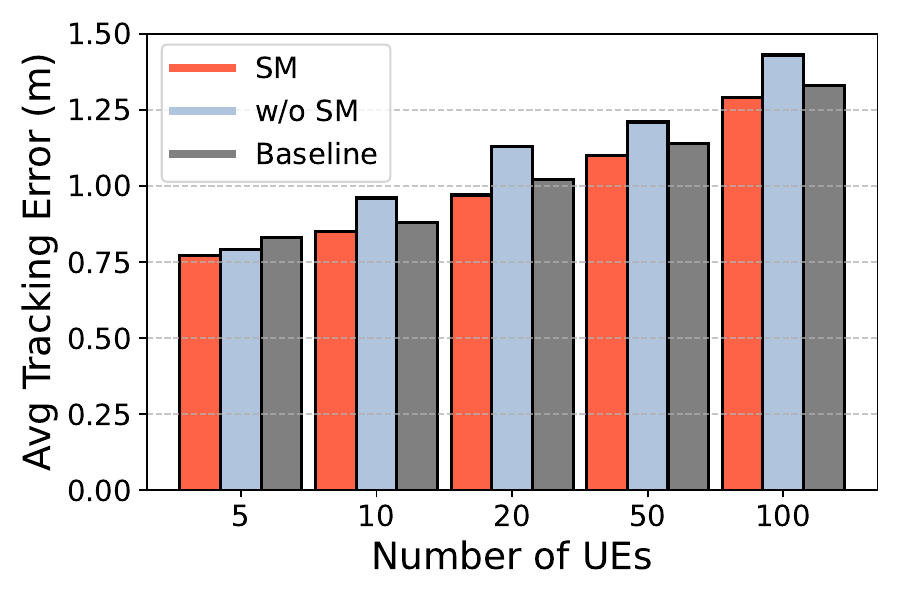}
        \caption{Performance with varying UE numbers}
        \label{fig:loc_error_ue}
    \end{subfigure}
    \hfill
    \begin{subfigure}[b]{0.24\textwidth}
        \centering
        \includegraphics[width=\linewidth]{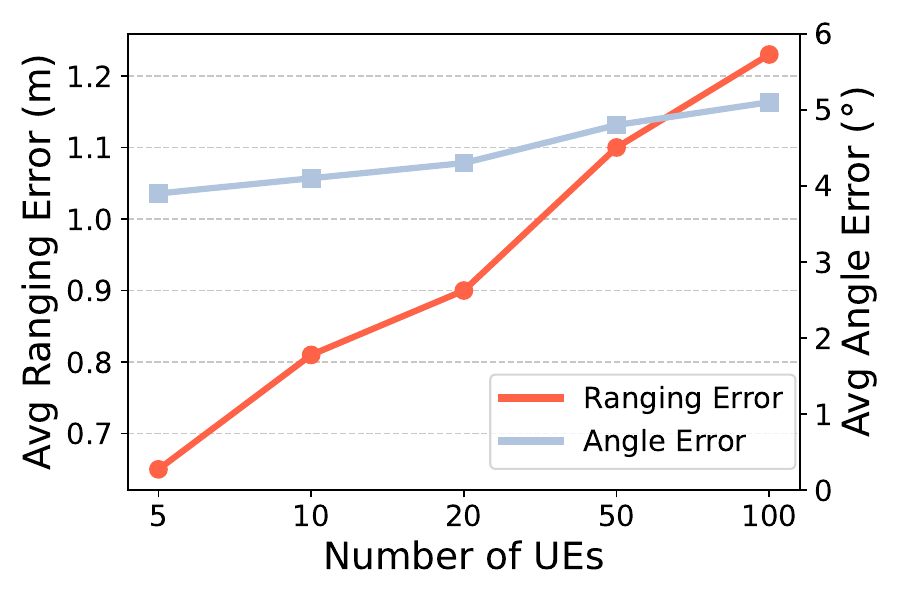}
        \caption{Tracking error decomposition analysis}
        \label{fig:ranging_angle_error}
    \end{subfigure}
    \caption{Sensing performance evaluation under various conditions.}
    \label{fig:sensing_performance}
\end{figure*}

Beyond communication performance, we evaluate the sensing capabilities of our proposed method in terms of positioning and tracking accuracy. Following the approach used in SpotiFi~\cite{kotaru2015spotfi}, we estimate the Angle of Arrival (AoA). However, instead of using Time of Flight (ToF), which has poor resolution, we rely on CSI amplitude to estimate the UE’s distance from the base station, as the coarse TA value is insufficient for precise ranging. Using these estimated AoA and distance values, we localize the UE and assess tracking accuracy by computing the Euclidean distance between estimated and true UE trajectories.

\subsubsection{Tracking Accuracy Under Different Traffic Loads}
We follow the same experimental setup as in the communication performance evaluation to assess the tracking accuracy of our adaptive SRS triggering algorithm across various traffic conditions. The effect of spatial smoothing (SM)  (\cref{sec:spatial_smooth})  is also evaluated in Fig.~\ref{fig:loc_error}. 
Tracking accuracy improves as traffic load increases, mirroring the trend observed in communication performance. SM plays a critical role in reducing tracking error, particularly in low-traffic conditions, where it lowers the average error by 12\% (from 1.11m to 0.97m). 
For comparison, we also evaluate a baseline using periodic SRS-only estimation, which already utilizes the full bandwidth and does not require spatial smoothing. Our adaptive scheme with spatial smoothing achieves \textbf{3.4\% lower tracking error} (0.85m vs. 0.88m) under high traffic conditions
and {11\% lower tracking error} (0.78m vs. 0.88m) under full traffic load.

\subsubsection{Tracking Accuracy in Different Environments}

Fig.~\ref{fig:loc_error_env} compares tracking accuracy across different environments. As expected, errors are slightly higher in the cluttered office than in the open-floor room, and periodic baseline also shows higher errors, consistent with the communication performance evaluation. Spatial smoothing significantly reduces tracking error in both environments, particularly in the cluttered office, where it lowers the average error by 14.5\% (from 0.96m to 0.82m). This shows that spatial smoothing effectively mitigates tracking errors caused by channel stitching in environments with strong multipath effects.

\subsubsection{Tracking Accuracy with Varying Numbers of UEs}

We also evaluate tracking accuracy under different numbers of active UEs (\( N \)) in high-traffic scenarios. As shown in Fig.~\ref{fig:loc_error_ue}, tracking error increases as \( N \) grows due to greater resource competition, leading to fewer sensing opportunities per UE.

However, spatial smoothing consistently reduces tracking error across all scenarios. The impact is especially pronounced as \( N \) increases where spatial smoothing reduces median tracking error by 2.5\% (0.79m to 0.77m) when $N = 5$ and by 14\% (1.13m to 0.97m) when $N = 20$.
This suggests that spatial smoothing effectively compensates for tracking errors introduced by resource competition among UEs.

Fig.~\ref{fig:loc_error_ue} also compares our method to periodic SRS-only estimation. As the number of UEs increases, periodic SRS leads to less frequent channel estimates, causing a rise in tracking error. Our adaptive scheme with spatial smoothing outperforms periodic SRS across all scenarios by weighting the most recent channel estimates more heavily in Eq.~\ref{eq:weighted_cov}.

Notably, our method with \( N = 10 \) achieves a tracking error of 0.85m—comparable to periodic SRS-only with \( N = 5 \) (0.83m), demonstrating that our approach can support more UEs while maintaining similar tracking accuracy.

\begin{figure*}[t!]
    \centering
    \begin{subfigure}[b]{0.3\textwidth}
        \centering
        \includegraphics[width=\linewidth]{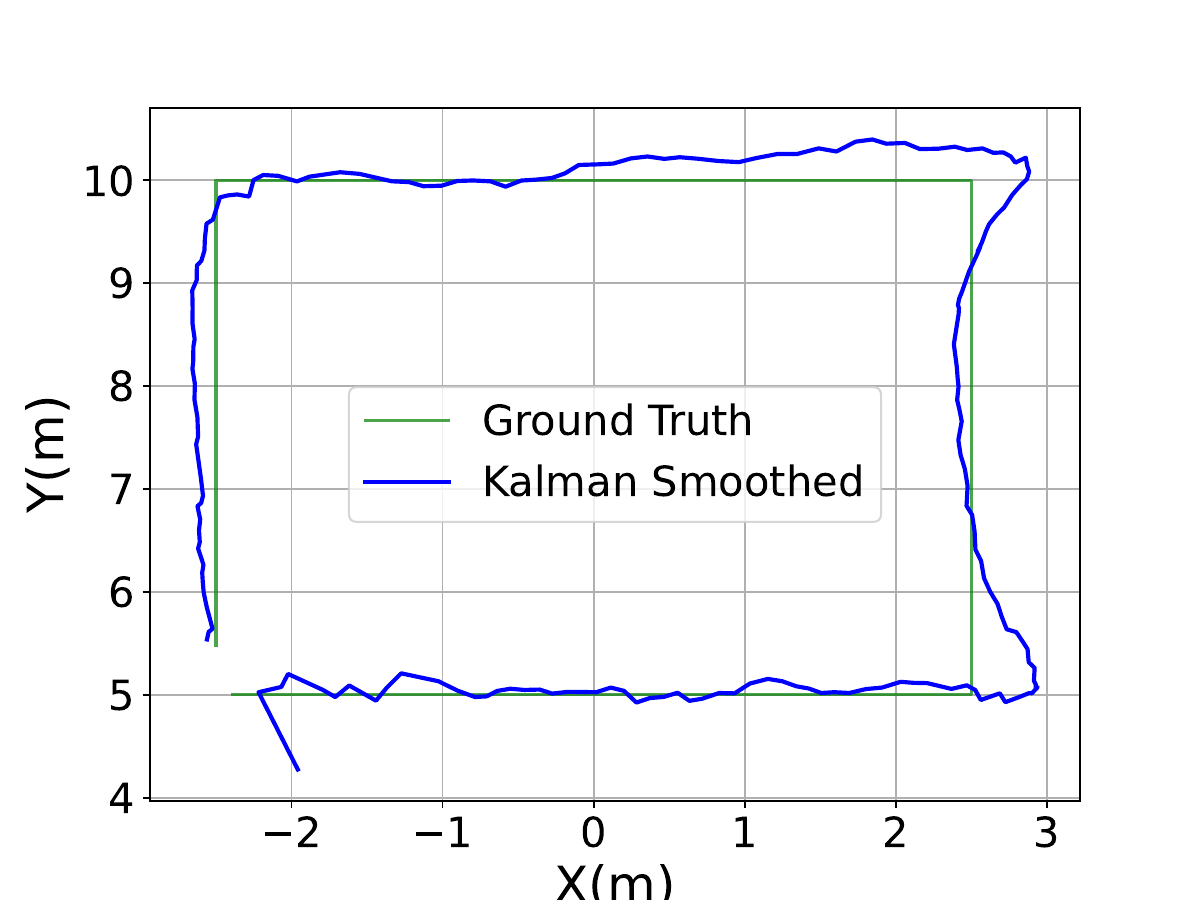}
        \caption{Rectangle}
        \label{fig:rec_traj}
    \end{subfigure}
    \hfill
    \begin{subfigure}[b]{0.3\textwidth}
        \centering
        \includegraphics[width=\linewidth]{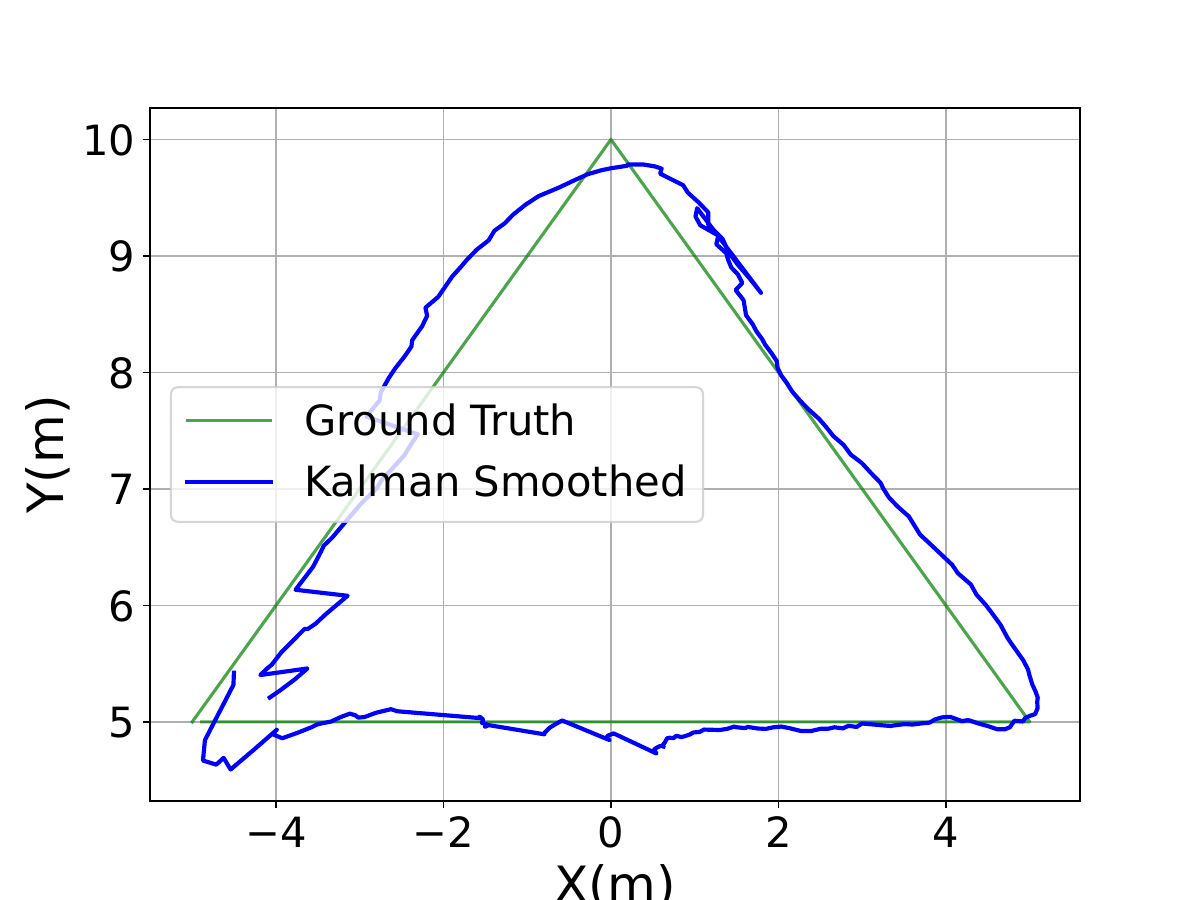}
        \caption{Triangle}
        \label{fig:tri_traj}
    \end{subfigure}
    \hfill
    \begin{subfigure}[b]{0.3\textwidth}
        \centering
        \includegraphics[width=\linewidth]{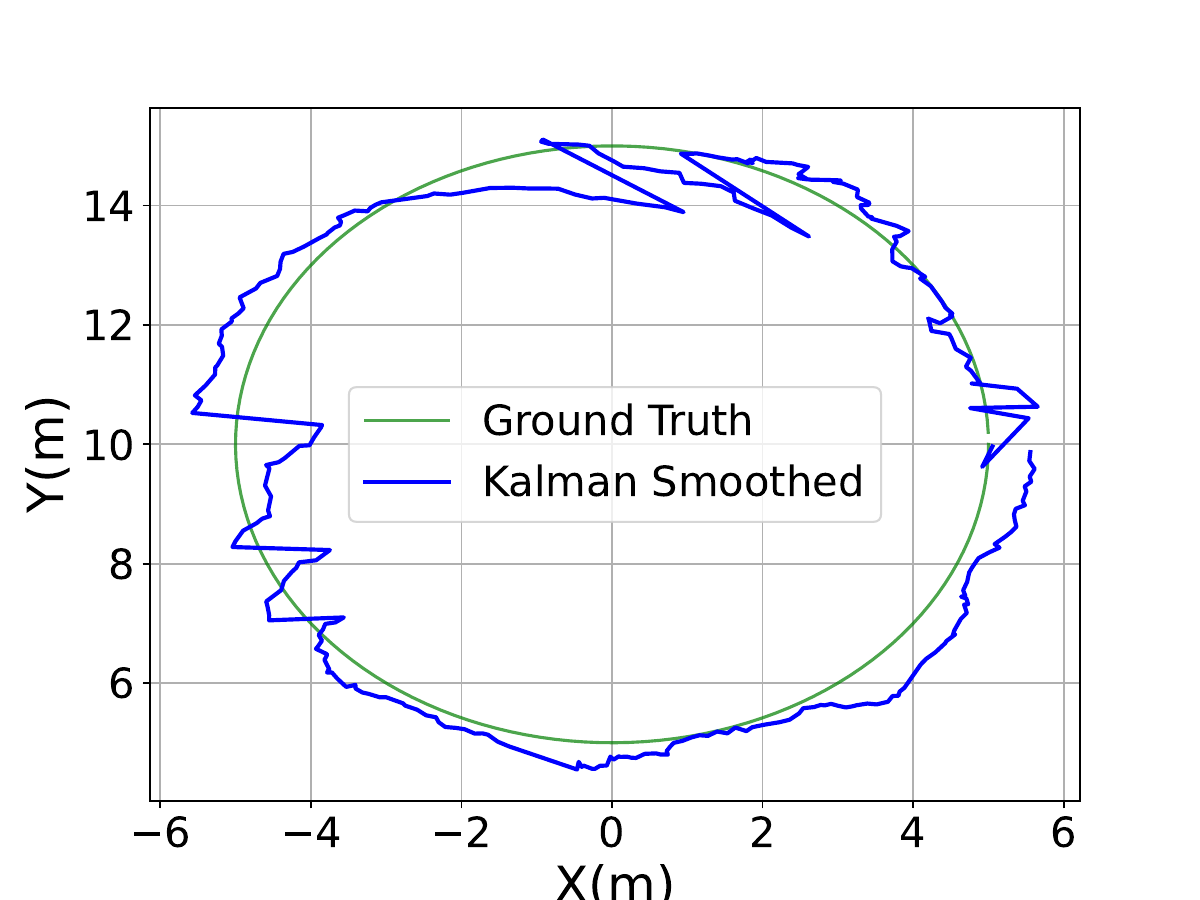}
        \caption{Circle}
        \label{fig:circ_traj}
    
    \end{subfigure}
        \vspace{-2mm}
    \caption{Estimated tracking trajectories with Kalman smoothing.}
    \label{fig:traj}
\end{figure*}

To further analyze tracking performance, we decompose the total tracking error into:
\begin{itemize}
    \item \textbf{Ranging error}, which depends on CSI amplitude and is highly sensitive to channel estimation errors.
    \item \textbf{Angular error}, which originates from AoA estimation inaccuracies.
\end{itemize}
Fig.~\ref{fig:ranging_angle_error} shows that, when the number of active UEs increases from 5 to 100, 
the ranging and angular errors increase from 0.65m to 1.23m (47\%) and from 3.9° to 5.1° (24\%).
These results indicate that ranging error is the primary contributor to tracking inaccuracies, particularly as the number of UEs increases, affecting system scalability. Since CSI amplitude-based ranging is more susceptible to channel estimation errors, future research could explore deep-learning-based approaches to enhance channel estimation accuracy and mitigate ranging errors. Finally, Fig.~\ref{fig:traj} shows that the Kalman filter smoothed trajectories closely align with the ground truth, \textcolor{re}{where the base station is positioned at the origin (0, 0) m}, demonstrating the effectiveness of our approach in maintaining tracking accuracy.

\section{DISCUSSION}

\textbf{Enhanced UE Capabilities.}  
While our current implementation demonstrates the feasibility of 5G sensing, significant opportunities exist to enhance UE capabilities within the current standard. For instance, increasing the maximum number of SRS resources per set and the maximum SRS resource sets could enable finer-grained SRS resource triggering. Additionally, future standards could introduce more flexible SRS allocation by allowing the DCI to dynamically assign SRS resources, similar to the PUSCH mechanism, thereby improving resource utilization efficiency.

\noindent\textbf{Fine-Grained Sensing Rate Adaptation.}  
Our current system employs a relatively coarse-grained approach to target sensing rate determination, whereas real-world environments require dynamic adaptation. Future work should explore context-aware sensing rate adaptation algorithms that balance sensing accuracy, energy consumption, and network resource utilization. These algorithms could leverage environmental factors (e.g., high-mobility vs. static scenarios), application requirements, and network conditions to optimize sensing parameters.

\noindent\textbf{Downlink Bistatic and Monostatic Sensing in 5G.}  
Our implementation primarily focuses on uplink sensing; however, a comprehensive sensing framework should integrate both downlink bistatic and monostatic sensing modalities. Bistatic sensing, where the transmitter and receiver are separated, offers distinct advantages in coverage and perspective diversity. Future research should develop coordination mechanisms between base stations to enable cooperative bistatic sensing, enhancing spatial resolution and coverage.

\textcolor{re}{\noindent\textbf{Passive Sensing.} Our work focuses on active sensing - localizing UEs, but potentially can be extended for passive sensing. The channel estimations generated from CARTS still include multipath reflections, which can be used to sense objects that are not transmitting. Two main challenges still exist. First, it is hard to separate objects' weak Doppler signals from the strong, direct signal of the moving UE. Second, sampling time and frequency offsets (STO and SFO) between the UE and gNB corrupt the signal's phase and ToF.}


\section{RELATED WORK}

\noindent\textbf{Wireless Sensing.}  
Wireless sensing has emerged as a promising technology that leverages wireless signals for contactless environmental and human activity sensing. Early work primarily utilized WiFi signals for motion detection and localization by analyzing CSI \cite{zhang2019widetect, jiang2014communicating, soltanaghaei2018multipath}. More recent research has explored finer-grained sensing capabilities, such as vital sign monitoring \cite{liu2015tracking, li2024spacebeat, hu2024m}, gesture recognition \cite{qian2017widar, abdelnasser2015wigest}, and human pose estimation \cite{ren2022gopose, jiang2020towards, ren20213d}.  
Advancements in this field have significantly improved sensing accuracy, enabling applications such as multi-person tracking \cite{karanam2019tracking}, sub-centimeter object profiling (WiProfile) \cite{yao2024wiprofile}, and robust through-wall 3D pose estimation (Wi-Vi) \cite{adib2013see}. 

\noindent\textbf{Wireless Channel Stitching.}  
Channel stitching techniques combine multiple frequency channels to create a wider effective bandwidth, enhancing both communication capacity and sensing resolution \cite{xiong15tonetrack, uvaydov2024stitching}. By coherently aggregating measurements across different frequency bands, these approaches overcome hardware bandwidth limitations and improve system performance.  
\textcolor{re}{Recent work has demonstrated efficient algorithms for bandwidth aggregation across heterogeneous spectrum chunks \cite{li2024uwb, pegoraro2024hisac, noschese2020multi, kazaz2021delay}; however, they assume either a static TX/RX setting or fairly large bandwidth (tens or hundreds MHz) to synchronize the CIR peaks. Compared with these approaches, though we only focus on aggregate contiguous bands, our method tackles challenges under moving TX, narrow band, and time-asynchronous conditions.}


\noindent\textbf{5G ISAC.}  
ISAC represents a paradigm shift in 5G and beyond, enabling communication infrastructure to simultaneously provide sensing capabilities \cite{zhang2021enabling}. This dual functionality offers significant advantages in spectrum efficiency, hardware reuse, and cost reduction compared to separate sensing and communication systems.  
Current research in 5G ISAC focuses on waveform design to balance sensing accuracy and communication performance \cite{liu2020joint}, as well as resource allocation strategies \cite{dong2022sensing}. Despite promising advancements, challenges remain in achieving optimal performance trade-offs, mitigating mutual interference, and deploying practical ISAC systems in dynamic environments \cite{chen2021radio}.

\section{CONCLUSION}
In this paper, we introduced \textbf{CARTS}, a novel framework that enhances 5G ISAC by fusing DMRS and SRS CSI measurements, addressing key challenges related to asynchronization, resource contention, and adaptive scheduling. By leveraging aperiodic SRS triggering, CARTS optimizes CSI acquisition without interfering with PUSCH scheduling while remaining fully compatible with existing 5G infrastructure. This makes it a scalable and cost-effective solution for next-generation wireless sensing and communication.  
Our evaluations demonstrate that CARTS improves CSI fidelity, reduces timing errors, and enhances UE tracking accuracy compared to periodic SRS-based schemes. We believe CARTS paves the way for a new paradigm in ISAC for 5G and beyond, truly integrating sensing and communication into a unified framework.


\balance
\bibliographystyle{ACM-Reference-Format}
\bibliography{reference}

\end{document}